\documentclass[acmsmall]{acmart}

\AtBeginDocument{%
  \providecommand\BibTeX{{%
    \normalfont B\kern-0.5em{\scshape i\kern-0.25em b}\kern-0.8em\TeX}}}

\setcopyright{acmcopyright}
\acmJournal{TOSEM}
\acmYear{2022} \acmVolume{1} \acmNumber{1} \acmArticle{1} \acmMonth{1} \acmPrice{15.00}\acmDOI{10.1145/3528100}

\usepackage{multirow}
\usepackage{subfig}
\usepackage{graphicx}
\usepackage{geometry}
\usepackage{bm}
\usepackage{xcolor}
\usepackage{amsthm}

\newtheorem{theorem}{Theorem}

\begin{document}

\title{HINNPerf: Hierarchical Interaction Neural Network for Performance Prediction of Configurable Systems}

\author{Jiezhu Cheng}
\orcid{0000-0002-1755-6828}
\affiliation{%
  \institution{Sun Yat-sen University}
  \streetaddress{No. 132, East of Outer Ring Road, Guangzhou Higher Education Mega Center, Panyu District}
  \city{Guangzhou}
  \state{Guangdong Province}
  \country{China}
  \postcode{510006}
}
\email{chengjzh@mail2.sysu.edu.cn}

\author{Cuiyun Gao}
\orcid{0000-0003-4774-2434}
\affiliation{%
  \institution{Harbin Institute of Technolgy, Shenzhen}
  \streetaddress{Xili University Town, Nanshan District}
  \city{Shenzhen}
  \state{Guangdong Province}
  \country{China}
  \postcode{518071}
}
\email{gaocuiyun@hit.edu.cn}

\author{Zibin Zheng}
\orcid{0000-0002-7878-4330}
\authornote{Corresponding author}
\affiliation{%
 \institution{Sun Yat-sen University}
 \streetaddress{No. 132, East of Outer Ring Road, Guangzhou Higher Education Mega Center, Panyu District}
 \city{Guangzhou}
 \state{Guangdong Province}
 \country{China}
 \postcode{510006}
}
\email{zhzibin@mail.sysu.edu.cn}

\renewcommand{\shortauthors}{J. Cheng et al.}

\begin{abstract}
Modern software systems are usually highly configurable, providing users with customized functionality through various configuration options. Understanding how system performance varies with different option combinations is important to determine optimal configurations that meet specific requirements. Due to the complex interactions among multiple options and the high cost of performance measurement under a huge configuration space, it is challenging to study how different configurations influence the system performance. To address these challenges, we propose \emph{HINNPerf}, a novel hierarchical interaction neural network for performance prediction of configurable systems. \emph{HINNPerf} employs the embedding method and hierarchic network blocks to model the complicated interplay between configuration options, which improves the prediction accuracy of the method. Besides, we devise a hierarchical regularization strategy to enhance the model robustness. Empirical results on 10 real-world configurable systems show that our method statistically significantly outperforms state-of-the-art approaches by achieving average 22.67\% improvement in prediction accuracy. In addition, combined with the Integrated Gradients method, the designed hierarchical architecture provides some insights about the interaction complexity and the significance of configuration options, which might help users and developers better understand how the configurable system works and efficiently identify significant options affecting the performance.
\end{abstract}



\maketitle

\section{Introduction}
\label{sec:Intro}

Modern configurable software systems offer customized services to users through a set of configuration options. By specifying a combination of these options (i.e., a \emph{configuration}), one can tailor the system's behavior to meet specific functional requirements. Meanwhile, the non-functional properties of the system may be significantly influenced by different configuration selections. Performance (such as response time and throughput) is one of the most important non-functional properties as it directly affects user experience and cost~\cite{DECART,DeepPerf}. It is necessary for both users and developers to efficiently identify the performance-optimal configurations when deploying and testing the software system. However, it is infeasible to exhaustively measure the system performance under all possible configurations, since the combinatorial explosion of configuration options results in an exponential number of configurations~\cite{SPLConquer_03,DECART,DeepPerf}.

Recently, researchers have devised various machine learning methods to predict system performance under any certain configuration~\cite{SPLConquer_01,SPLConquer_02,SPLConquer_03,CART,DECART,Fourier,PerLasso,DeepPerf}. In this way, a \emph{performance model} is constructed to ease understanding, debugging, and optimization of highly configurable software systems. For example, the performance model can help a user find the best performing configuration under specific functional constraints. And a developer may compare the performance model with his own mental model to check whether the software behaves as expected, and improve the software based on the comparison~\cite{SPLConquer_03}. Compared to the huge time cost of measuring the system performance by executing a complex benchmark, a performance model can predict the performance under a certain configuration within a few seconds, which significanlty reduces software testing cost.

To build a performance model, the scalar performance values of a set of configurations are required to be first measured. The configurations and their corresponding performance measurements compose a \emph{sample} for training a model which later can be used for predicting the performance values of new configurations. Most importantly, prediction accuracy determines whether a performance model can really help users and developers improve software quality, rather than causing unnecessary misleading. This is why most of prior work have focused on improving the accuracy of performance models~\cite{SPLConquer_01,SPLConquer_02,SPLConquer_03,CART,DECART,Fourier,PerLasso,DeepPerf}. Nevertheless, learning a performance model with high accuracy is very challenging because of the following aspects:
\begin{enumerate}
  \item \textbf{Complex interactions among features}: Configurable systems typically have multiple binary and/or numeric configuration options (also called \emph{features}). The interactions among different features can be non-linear, multi-way and hierarchical~\cite{SPLConquer_02,HierarchicalInter}. Failing to learn the interactions that contribute substantially to software performance causes significant drop on prediction accuracy.
  \item \textbf{Small sample}: Typically, measuring the performance of a configuration requires executing a realistic workload on the whole system, which is costly and time-consuming~\cite{PerfHistory}. Hence, only a limited set of configurations can be measured as a training sample in practice~\cite{DECART}. A small sample could cause the performance model to easily overfit (i.e., the model performs well on the training data but badly during testing).
\end{enumerate}

Existing methods tackle the above challenges in varied ways. For interaction modeling, \emph{SPLConqueror}~\cite{SPLConquer_01,SPLConquer_02,SPLConquer_03} learns the influences of individual configuration options (features) and their interplay on performance by multiple linear regression. However, the method only models the interactions with linear combinations of low-order functions (e.g., linear, quadratic and logarithmic),  limiting its flexibility in recognizing complex non-linear interactions. Further, \emph{CART/DECART}~\cite{CART,DECART} employs classification and regression trees to build a more flexible non-linear performance model. Fourier learning algorithms~\cite{Fourier,PerLasso} transform the regression problem into its Fourier form and predict the performance by estimating the Fourier coefficients. The main disadvantage of \emph{CART/DECART} and Fourier methods is that they can only learn the interactions among binary configuration options. \emph{DeepPerf}~\cite{DeepPerf} introduces a deep feedforward neural network (FNN) to overcome the shortcomings of previous methods. Nonetheless, a deep FNN can only model the feature interactions implicitly through multiple hidden layers, which probably limits its flexibility and accuracy. For example, some hierarchical configuration interactions prone to affecting the system performance are hard to be engaged to \emph{DeepPerf}, since the interactions between different layers of FNN is opaque~\cite{FNN_hier}. On the other hand, to mitigate the limitation of the small sample, both \emph{SPLConqueror}~\cite{SPLConquer_03} and \emph{DECART}~\cite{DECART} incorporate several sampling heuristics to select a set of representative configurations for training. However, it takes extra time and effort to determine the appropriate sampling strategy for each system since there is no universal optimal sampling heuristics~\cite{bullet}. Instead, \emph{PerLasso}~\cite{PerLasso} and \emph{DeepPerf}~\cite{DeepPerf} focus on restricting the performance model to be sparse with $L_1$ regularization. This strategy takes the risk of making the model highly sensitive to the regularization hyperparameter, that is, small changes in hyperparameter may cause large fluctuations in model accuracy. In this case, the regularization hyperparameter must be carefully tuned in a large search space, which increases model training cost.

To better address the aforementioned challenges, we propose a novel \underline{H}ierarchical \underline{I}nteraction \underline{N}eural \underline{N}etwork for \underline{Perf}ormance Prediction and name it \emph{HINNPerf}. Similar to \emph{DeepPerf}~\cite{DeepPerf}, our method can model all types of complex interactions (i.e., binary, numeric, and binary-numeric interactions) in various configurable systems. Differently, inspired by previous work~\cite{SPLConquer_02,HierarchicalInter,SPLConquer_03} revealing that interactions among configuration options are hierarchical, we decompose the deep neural network architecture into multiple \emph{blocks} connected hierarchically, where lower (higher) blocks learn the influences of lower-order (higher-order) interactions on system performance. Within each block, similarly to embedding methods~\cite{FNN_embed_01,FNN_embed_02,FNN_embed_04} widely used in deep learning world, we employ an FNN to embed the feature interactions into a vector. All the blocks link to each other by the vector concatenation. Also, each block predicts a partial performance value and the final prediction is the sum of all partial predictions. In this way, it is much easier for \emph{HINNPerf} to learn the influences of individual features and their interactions on performance via a deeper and more \textit{expressive} architecture\footnote{Here we refer the \textit{expressive} power of a method to the capability of learning various functions~\cite{expressive}. More expressive methods can learn more complex functions.}. In addition, based on the sparsity prior of the software performance functions~\cite{SPLConquer_02,SPLConquer_03,DeepPerf}, we engage the $L_1$ regularization~\cite{lasso} for the first FNN layer of each block to address the challenge of small sample. We empirically find that such hierarchical regularization architecture is more robust to the regularization hyperparameter and reduces hyperparameter tuning cost. As for practicality, on the one hand, the hierarchical structure of \emph{HINNPerf} enables it to provide some insights on configuration coupling of the system, which can help users and developers understand which kind of interactions (lower- or higher-order) has the major influence on the system performance. On the other hand, combined with the Integrated Gradients~\cite{IG} method, \emph{HINNPerf} recognizes the significant configuration options in different interaction orders and reveals some potential interaction patterns for the system, presenting more actionable information for users and developers. Finally, it can be demonstrated that the \emph{DeepPerf}~\cite{DeepPerf} model is simply a special case of our \emph{HINNPerf} architecture. From this perspective, our method provides a general deep learning framework for performance prediction of configurable systems.

We evaluate our method on 10 real-world configurable software systems including compilers, web servers, video encoders, etc. The experimental results show that \emph{HINNPerf} outperforms the state-of-the-art methods on most systems. Remarkably, when compared to the advanced deep learning method \emph{DeepPerf}~\cite{DeepPerf}, \emph{HINNPerf} achieves statistically significant improvements on binary-numeric systems (i.e., the systems with both binary and numeric configuration options), demonstrating its strength in capturing complex feature interactions.

In summary, our contributions are of four-folds:
\begin{itemize}
  \item We introduce the embedding method and propose a novel deep neural network architecture to model the hierarchical interactions among configuration options and predict the performance of configurable software systems. To our knowledge, although the embedding method has achieved great success in various deep learning tasks~\cite{FNN_embed_01,FNN_embed_02,FNN_embed_03,FNN_embed_04}, few literatures have employed it to model the feature interactions of configurable systems. And we aim to bridge the gap in this work.
  \item We devise a hierarchical regularization strategy to help the $L_1$ regularization address the small sample challenge more efficiently, with the advantages of improving model robustness to the regularization hyperparameter and reducing hyperparameter tuning cost.
  \item We conduct extensive experiments on 10 real-world configurable software systems with various sample sizes and show the advantages of our method against most state-of-the-art baseline approaches, demonstrating new benchmark on the public datasets. Our experimental data and source code are publicly available at the Google Drive\footnote{\url{https://drive.google.com/drive/folders/1qxYzd5Om0HE1rK0syYQsTPhTQEBjghLh?usp=sharing}}.
  \item In addition to accuracy improvement, our method provides some insights on the complexity of configuration coupling and identifies the significant configuration options in different interaction orders by automatic learning, which could help users and developers better understand how the configurable system works.
\end{itemize}

\section{Background and Motivation}
\label{sec:Motivation}

\subsection{Problem Formulation}

Table~\ref{tab:con_perf} illustrates a performance example of a configurable software system. The system has $14$ configuration options with $11$ binary options and $3$ numeric options. Each row of the table represents one configuration and its performance measurement that can be response time, throughput, workload and so on. Evidently, different configurations usually lead to different performance values. Our main task is to predict the performance value of a new configuration that is not measured (e.g., the value $?$ in the last row of Table~\ref{tab:con_perf}), based on the observations of previously measured configurations.

Formally, we aim to learn a function that maps a configuration with $n$ options $\mathbf{o} = [o_1, o_2, ..., o_n]^\top$ to its performance value:
\begin{equation}
  p = f(\mathbf{o}) = f(o_1, o_2, ..., o_n), \label{equ:perf_func}
\end{equation}
where $f: \mathbb{X} \to \mathbb{R}$ is the performance function and $\mathbb{X}$ is the Cartesian product of the domains of all the configuration options. Naively, we can derive a precise performance function by exhaustively measuring the performance of every valid configuration of the software system, which is infeasible in practice because of the exponentially growing configuration space~\cite{SPLConquer_03,DECART,DeepPerf,config_space}. Therefore, the objective is to design a performance model for precisely approximating the performance function using only a small sample.

\begin{table}[htbp]
  \caption{Performance values of different configurations}
  \label{tab:con_perf}
  \begin{tabular}{|c|c|c|c|c|c|c|c|}
    \cline{1-8}
    \multicolumn{7}{|c|}{Configuration Options} & Performance \\
    \cline{1-8}
    $o_1$ & $o_2$ & $o_3$ & ... & $o_{12}$ & $o_{13}$ & $o_{14}$ & $p$ \\
    \cline{1-8}
    $0$   & $1$   & $1$   & ... & $64$     & $1$      & $0$      & $101.754$ \\
    $0$   & $1$   & $0$   & ... & $64$     & $5$      & $6$      & $567.883$ \\
    $1$   & $1$   & $0$   & ... & $256$    & $2$      & $3$      & $219.512$ \\
    .     & .     & .     & ... & .        & .        & .        & .         \\
    $0$   & $1$   & $0$   & ... & $4096$   & $5$      & $4$      & $399.705$ \\
    $0$   & $1$   & $0$   & ... & $4096$   & $6$      & $6$      & $643.311$ \\
    $0$   & $1$   & $0$   & ... & $256$    & $3$      & $3$      & $?$       \\
    \cline{1-8}
\end{tabular}
\end{table}

\subsection{Hierarchical Interactions}
\label{subsec:hier_inter}

A large number of configuration options with different types (binary and numeric) can generate complex interactions that affect configurable system performance in different ways. In theory, any combination of the configuration options may cause a distinct interaction pattern~\cite{SPLConquer_02}. Fortunately, researchers have found that relevant interactions form a hierarchical relationship~\cite{SPLConquer_02,HierarchicalInter,SPLConquer_03}. That is, higher-order interactions usually build on lower-order interactions. As is pointed out by~\cite{SPLConquer_03}, three-way interactions (i.e., interactions among three options) build on corresponding two-way interactions among the same set of options. To improve the accuracy of performance prediction, it is necessary to model the influence of diverse interaction patterns on performance through a hierarchical and expressive method.

\subsection{Deep Feedforward Neural Network}
\label{subsec:deep_res_net}

Deep neural networks are one of the most expressive models owing to their excellent strength in capturing non-linear data relationships~\cite{DeepL}. A feedforward neural network (FNN) is a network which connects the input and the output through multiple stacked layers (hidden layers) of computational units (neurons)~\cite{DeepL,DeepPerf}. According to the universal approximation theorem~\cite{approx_01,approx_02,approx_03}, an FNN with at least one hidden layer can approximate any continuous function from one finite dimensional space to the other at any level of accuracy, as long as enough neurons and a suitable activation function (e.g., sigmoid, ReLU) are provided~\cite{DeepPerf}. Hence, it is feasible to approximate the performance prediction function of a configurable system with an FNN. However, the number of neurons needed to approximate a real-world function might be infinitely large and it is impossible to train such an FNN in finite time. Instead, researchers have tuned to train deep FNNs under an upper bound of the approximation error~\cite{why_deep,error_bound}. Theoretically, a deeper FNN can better enrich the "levels" of data features and is more expressive to approximate functions with higher level of accuracy~\cite{residualNN,deep_power_01,deep_power_02}. Nevertheless, we propose that a deep FNN is probably not a sufficiently good architecture for the performance prediction problem, mainly due to following reasons:
\begin{itemize}
	\item The interaction modeling process of a deep FNN is opaque. Under a general FNN architecture, features of lower hidden layers serve as inputs to higher hidden layers and the final output is the non-linear combinations of the features in the last hidden layer. In this way, it is easy for a deep FNN to learn a performance function of high-level data features but ignore the hierarchical interactions between lower- and higher-order features. Such a limitation might not be well suited for the hierarchical interaction nature of configurable software systems and reduce the prediction accuracy.
	\item The FNN architecture is a complete black box model such that it can only predict the scalar performance value without providing insights about how feature interactions of different orders influence the system performance.
\end{itemize}

To overcome above limitations of FNN, deep learning researchers have extended general FNNs into hierarchical network architectures, achieving significant improvements in image processing~\cite{residualNN,hier_img}, natural language processing~\cite{FNN_embed_01,hier_nlp}, time series prediction~\cite{NBEATS}, etc. Similarly in this paper, we devise a novel network architecture consisting of multiple blocks connected hierarchically, where different network blocks learn the influence of different interactions on system performance. Our method well fits the hierarchical interaction nature of configurable systems and thus achieves higher accuracy than the \emph{DeepPerf}~\cite{DeepPerf} method. Besides, by measuring the contributions of different blocks to the system performance and computing the significance score of each configuration option in different interaction orders, our approach provides some additional insights into the coupling complexity and interaction patterns of the system, as we will discuss in later sections.

\subsection{Interaction Embedding}
\label{subsec:inter_embed}

The main task of the embedding method is to translate an object from a high-dimensional abstract space (e.g., a word, an image, a video, etc) into a vector representation of a low-dimensional space, while maintaining some basic properties such as semantics and temporal relationships. Various embedding methods have been proposed to improve a variety of deep learning tasks such as natural language processing~\cite{FNN_embed_01,FNN_embed_05,FNN_embed_06}, recommendation systems~\cite{FNN_embed_02,FNN_embed_07}, and video encoding\cite{FNN_embed_04}.

As we described in Section~\ref{sec:Intro}, the interaction space of configurable systems could be extremely high-dimensional due to the exponential configuration space and complex interaction patterns. Hence, it is reasonable to embed the interactions into low-dimensional vectors and model their influence on the performance. However, there is little work about introducing embedding methods into performance prediction of configurable systems and we hope to make a step forward. Following the popular neural network language embedding method~\cite{FNN_embed_01}, we employ multiple FNNs to perform interaction embedding of different orders and empirically find it beneficial for prediction accuracy improvement.

\subsection{Sparsity Regularization}
\label{subsec:sparse_reg}

In general, training a deep neural network with high prediction accuracy requires large-scale datasets, which is not applicable for the performance prediction task domain. A good way to mitigate the limited dataset issue is to incorporate prior knowledge into the neural network to guide its learning process~\cite{DeepPerf}. One important prior knowledge is that only a small number of configuration options and their interactions have a significant impact on system performance~\cite{SPLConquer_02,SPLConquer_03,DeepPerf}, implying sparsity on the parameters of the performance model (i.e., making parameters of many insignificant options and interactions equal to zero).

The $L_1$ regularization~\cite{lasso} is one of the most widely-used techniques for sparsity enforcement. By adding a penalty term constructed from the sum of the absolute values of model parameters to the loss function, $L_1$ regularization encourages the magnitude of parameters to be small and even zero~\cite{L1Reg}. However, the model accuracy might be highly sensitive to the regularization hyperparameter and it usually takes a lot of effort and time to search for optimal value on a large hyperparameter space~\cite{DeepPerf,PerLasso}. To reduce training cost, it is necessary to design an effective mechanism to improve model robustness and reduce the hyperparameter space.

\section{Approach}
\label{sec:approach}

We propose \emph{HINNPerf}, a hierarchical interaction neural network to predict performance values of configurable systems. Based on the discussion in Section~\ref{sec:Motivation}, our network architecture design relies on three key principles. First, for better interaction modeling, the architecture should incorporate the embedding method and the hierarchical interaction nature of configurable systems. Second, considering the small sample challenge, the architecture should be sparse but efficient for hyperparameter tuning. Third, the architecture should provide some insights for users and developers on how feature interactions of different orders affect the system performance. We now describe in detail how these principles converge to our method.

\subsection{Theoretical Model}
\label{subsec:theo_model}

We construct the theoretical model by decomposing the performance function $f$ of Equation~(\ref{equ:perf_func}) into the sum of $m$ subfunctions:
\begin{align}
\begin{split}
  &f(\mathbf{o}) = \sum_{j=1}^m f_j(\mathbf{x}_j)~, \\
  \mathbf{x}_j =& \begin{cases}
    ~~~\mathbf{o} & j = 1 \\
    \mathbf{x}_{j-1} \oplus I_{j-1}(\mathbf{x}_{j-1}) & 2 \le j \le m
  \end{cases},
\end{split} \label{equ:perf_decom}
\end{align}
where $\mathbf{o} \in \mathbb{X} \subset \mathbb{R}^n$ is the configuration with $n$ options and $\oplus$ denotes the vector concatenation operation. The model defined by Equation~(\ref{equ:perf_decom}) contains two function families: 
\begin{itemize}
  \item $\{I_j: \mathbb{R}^{jn} \to \mathbb{R}^n\}_{1 \le j \le m}$, the interaction embedding functions. Each $I_j$ learns the interactions among different features of input $\mathbf{x}_j$ and embeds the results into a new vector. Here we set the embedding dimension to $n$, the same as the number of options in configuration $\mathbf{o}$. That is, each $I_j$ embeds the interaction information into a vector of $\mathbb{R}^n$ space. With this condition, we can infer that $\mathbf{x}_j \in \mathbb{R}^{jn}$ from the definition of Equation~(\ref{equ:perf_decom}). Note that the embedding dimension could be designed flexibly according to the demands of the real-world environmental settings.
  \item $\{f_j: \mathbb{R}^{jn} \to \mathbb{R}\}_{1 \le j \le m}$, the performance subfunctions. Each $f_j$ maps the input features $\mathbf{x}_j \in \mathbb{R}^{jn}$ to a partial performance value, which is summed with other partial values into the actual performance value $f(\mathbf{o})$.
\end{itemize}

The design rationale of our theoretical model is exactly the hierarchical interaction rule described in Section~\ref{subsec:hier_inter} and the interaction embedding described in Section~\ref{subsec:inter_embed}. In this case, the higher-order interaction $\mathbf{x}_j$ is built by the concatenation of lower-order interactions $\mathbf{x}_{j-1}$ and $I_{j-1}(\mathbf{x}_{j-1})$, which facilitates the performance representation of different interaction orders. For example, subfunction $f_1$ might only represent the partial performance value under the influences of original configurable options and the two-way interactions among them. And the embedding function $I_1$ might embed these two-way interactions into the vector $I_1(\mathbf{o})$. By concatenating the original configuration $\mathbf{o}$ with its interaction embedding vector $I_1(\mathbf{o})$ as the input for $f_2$, it is easier for $f_2$ to represent the partial performance value under the influences of three-way interactions. Similarly, higher-order subfunctions $f_j (j \ge 3)$ correspond to performance influences of higher-order interactions. Note that the interaction patterns of real-world configurable software may be much more complex than the simple example above, but our theoretical model is flexible to handle different situations. 

\subsection{Network Architecture}

\begin{figure*}[!t]
  \centering
  \includegraphics[width=0.9\linewidth]{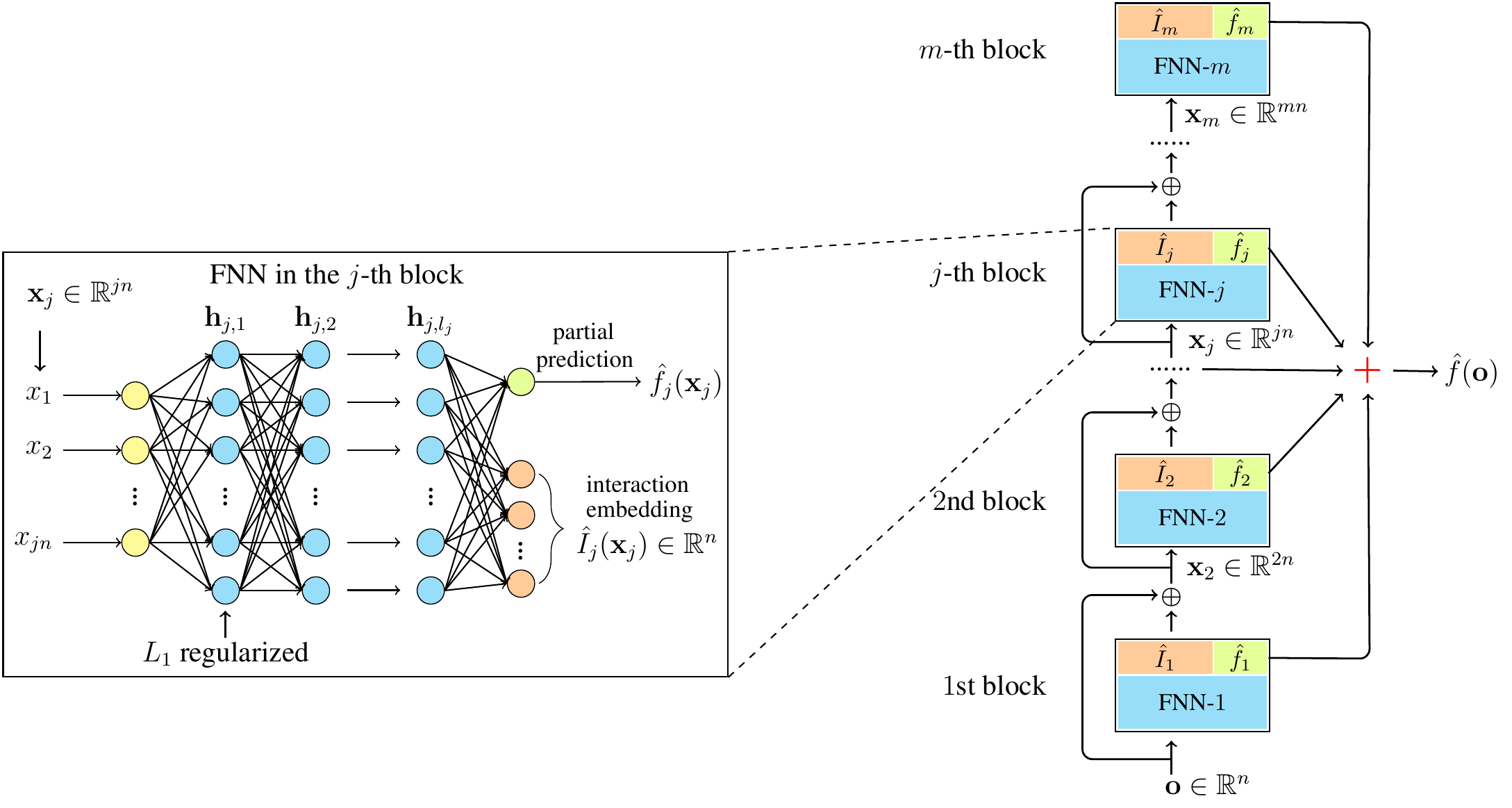}
  \caption{An overview of the proposed \emph{HINNPerf} architecture.}
  \Description{building block of \emph{HINNPerf}.}
  \label{fig:block}
\end{figure*}

According to our theoretical model, we design a hierarchical interaction network architecture to approximate all the performance subfunctions and interaction embedding functions in Equation~(\ref{equ:perf_decom}). Figure~\ref{fig:block} (right) presents an overview of the proposed \emph{HINNPerf} architecture. Similar in spirit to the \emph{N-BEATS} model~\cite{NBEATS} and other hierarchical network architectures~\cite{hier_img,hier_nlp}, our model consists of multiple hierarchically connected building blocks. Figure~\ref{fig:block} (left) shows the $j$-th block of the \emph{HINNPerf} architecture. The $j$-th block accepts its input $\mathbf{x}_j$ and employs a multi-layer feedforward neural network (FNN) to approximate subfunction $f_j$ and embedding function $I_j$. The input and the hidden layers of FNN are shared by the two approximation functions $\hat{f}_j$ and $\hat{I}_j$ while different output layers are utilized to produce partial performance prediction $\hat{f}_j(\mathbf{x}_j)$ and interaction embedding vector $\hat{I}_j(\mathbf{x}_j)$, respectively. 

Suppose the FNN of the $j$-th block has $l_j$ hidden layers with each layer containing $d_j$ neurons, the computation of the $j$-th block is described by the following equations:
\begin{align}
\begin{split}
\mathbf{h}_{j, 1} &= \text{ReLU}(\mathbf{W}_{j,1}^\top \mathbf{x}_j + \mathbf{b}_{j,1}), \\
\mathbf{h}_{j, 2} &= \text{ReLU}(\mathbf{W}_{j,2}^\top \mathbf{h}_{j, 1} + \mathbf{b}_{j,2}), \\ 
&\text{......}~, \\
\mathbf{h}_{j, l_j} &= \text{ReLU}(\mathbf{W}_{j,l_j}^\top \mathbf{h}_{j, l_j-1} + \mathbf{b}_{j,l_j}), \\
\hat{f}_j(\mathbf{x}_j) &= \mathbf{W}_{j,f}^\top \mathbf{h}_{j, l_j} + b_{j,f}, \\
\hat{I}_j(\mathbf{x}_j) &= \text{ReLU}(\mathbf{W}_{j,I}^\top \mathbf{h}_{j, l_j} + \mathbf{b}_{j,I}),
\end{split} \label{equ:multi_block}
\end{align}
where:
\begin{itemize}
  \item $\mathbf{W}_{j,k}, \mathbf{b}_{j,k} (1 \le k \le l_j)$ are weights and biases of the $k$-th hidden layer.
  \item ReLU is the activation function for non-linear learning in deep FNN~\cite{ReLU_1,ReLU_2}.
  \item $\mathbf{h}_{j,k} \in \mathbb{R}^{d_j} (1 \le k \le l_j)$ is the output vector of the $k$-th hidden layer.
  \item $\mathbf{W}_{j,f} \in \mathbb{R}^{d_j}, b_{j,f} \in \mathbb{R}$ are weights and bias used for predicting the partial performance value $\hat{f}_j(\mathbf{x}_j)$, respectively. Note that $\hat{f}_j(\mathbf{x}_j)$ is the linear output of $\mathbf{h}_{j,l_j}$ since the performance prediction is a regression problem~\cite{DeepPerf}.
  \item $\mathbf{W}_{j,I} \in \mathbb{R}^{d_j \times n}, \mathbf{b}_{j,I} \in \mathbb{R}^n$ are weights and biases that project the output vector of the last hidden layer to the interaction embedding space, respectively.
\end{itemize}

\subsection{Hierarchical Regularization}

As discussed in Section~\ref{sec:Intro} and Section~\ref{subsec:sparse_reg}, training a deep neural network on the small sample of a configurable system is difficult because of the high risk of overfitting. One way to solve the problem is to introduce additional restriction to the network parameters by employing a suitable regularization technique. Instructively, researchers have found that for configurable software systems, although there are exponential number of interactions among configuration options, a very large portion of potential interactions has no influence on system performance~\cite{SPLConquer_02,SPLConquer_03,DeepPerf}. This observation inspires us that parameters corresponding to unimportant interactions should be eliminated in the performance model, that is, parameters of the performance model could be quite sparse. Under the sparsity condition, we decide to apply the $L_1$ regularization technique to our method because of its strength on sparsity enforcement. 

Denote $\bm{\theta}$ as the parameters of a performance model, $\mathbf{X}$ as the input data, $\mathbf{Y}$ as the output data and $J(\bm{\theta},\mathbf{X},\mathbf{Y})$ as the loss function of the model. Generally, The $L_1$ regularization changes the loss function from $J(\bm{\theta},\mathbf{X},\mathbf{Y})$ to:
\begin{equation}
J_{\text{reg}}(\bm{\theta},\mathbf{X},\mathbf{Y}) = J(\bm{\theta},\mathbf{X},\mathbf{Y}) + \lambda ||\bm{\theta}||_1,
\end{equation}
where $||\cdot||_1$ denotes the $L_1$ norm and $\lambda$ is the regularization hyperparameter. For a deep FNN, $\bm{\theta}$ includes weights and biases of all the hidden layers in the network. However, as discussed in~\cite{DeepPerf}, it is infeasible to apply regularization to all the hidden layers in deep FNN due to the diversity and high sensitivity of the regularization hyperparameters in different hidden layers. Hence, \cite{DeepPerf}~only applies $L_1$ regularization to the first hidden layer of their \emph{DeepPerf} model, which achieves a great improvement on performance prediction. The disadvantage of such strategy is that the sparsity of the whole network is controlled only by the regularization of the first hidden layer. In this case, the behavior of the model heavily depends on the regularization hyperparameter of the first hidden layer and even small changes of $\lambda$ may cause large fluctuations in prediction accuracy. Therefore, one should carefully search for the optimal hyperparameter value on a large numerical space in order to achieve desired results, which increases the cost of hyperparameter tuning.

Taking the advantage of the hierarchical network architecture of our \emph{HINNPerf} model, we suggest to apply $L_1$ regularization to the first layer of the FNN in each block with one global regularization hyperparameter, as shown in Figure~\ref{fig:block} (left). The loss function of our method is changed to:
\begin{equation}
J_{\text{reg}}(\bm{\theta},\mathbf{X},\mathbf{Y}) = J(\bm{\theta},\mathbf{X},\mathbf{Y}) + \lambda \sum_{j=1}^m (||\mathbf{W}_{j,1}||_1 + ||\mathbf{b}_{j,1}||_1). \label{equ:L_1_lambda}
\end{equation}
Such hierarchical regularization technique allows us to shrink the parameters of unnecessary interactions in different orders and thus is more flexible and effective. In our experiments, we show that our method is more robust to the regularization hyperparameter and our hyperparameter search space is reduced by 6 times compared to the \emph{DeepPerf} model~\cite{DeepPerf}.

\subsection{Complexity Analysis}
\label{subsec:complexity}

We now analyze the complexity of our \emph{HINNPerf} model through the scale of network parameters. For model construction, we set the number of hidden layers and the number of hidden neurons to be the same in all blocks, that is, we set $l_1 = l_2 = \cdots = l_m = l$ and $d_1 = d_2 = \cdots = d_m = d$. According to Equation~(\ref{equ:multi_block}), the $j$-th block of \emph{HINNPerf} has $(jn+1)d + (l-1)(d+1)d + (d+1)(n+1)$ parameters, where:
\begin{itemize}
\item the first term $(jn+1)d$ comes from the weights $\mathbf{W}_{j,1} \in \mathbb{R}^{jn \times d}$ and biases $\mathbf{b}_{j,1} \in \mathbb{R}^{d}$ of the first hidden layer; 
\item the second term $(l-1)(d+1)d$ corresponds to the weights $\mathbf{W}_{j,k} \in \mathbb{R}^{d \times d}$ and biases $\mathbf{b}_{j,k} \in \mathbb{R}^{d}$ ($2 \le k \le l$) of the last $l-1$ hidden layers;
\item the third term $(d+1)(n+1)$ is produced by the weights and biases of the interaction embedding ($\mathbf{W}_{j,I} \in \mathbb{R}^{d \times n}, \mathbf{b}_{j,I} \in \mathbb{R}^n$) and the partial performance output ($\mathbf{W}_{j,f} \in \mathbb{R}^{d}, b_{j,f} \in \mathbb{R}$).
\end{itemize}
Hence, the scale of network parameters in the $j$-th block is $\mathcal{O}(jnd+ld^2)$, implying total $\mathcal{O}(m^2nd+mld^2)$ complexity of the \emph{HINNPerf} model. On the other hand, suppose that the \emph{DeepPerf}~\cite{DeepPerf} model has $l'$ hidden layers and the same $d$ neurons in each hidden layer as \emph{HINNPerf}, the total complexity of \emph{DeepPerf} is $\mathcal{O}(nd + l'd^2)$. Note that $l < l'$ since the number of hidden layers in each block of \emph{HINNPerf} is typically smaller than the total number of hidden layers in \emph{DeepPerf}, and we roughly have $ml \approx l'$. Accordingly, \emph{HINNPerf} mainly increases $\mathcal{O}(m^2nd)$ complexity compared to \emph{DeepPerf}. However, based on the previous works~\cite{SPLConquer_02,SPLConquer_03,HierarchicalInter} revealing that there are at most \emph{five-way} interactions in configurable systems, we restrict $m \le 5$ in our model. Therefore, the increase in model complexity caused by the parameter $m$ could be regarded as constant (i.e., $\mathcal{O}(m^2nd) \approx \mathcal{O}(Mnd)$ where $M$ is a constant).

Another question is that how many configuration options can be handled by our model? Is there an upperbound of the option number $n$ for \emph{HINNPerf}? We state that according to the the universal approximation theorem~\cite{approx_01,approx_02,approx_03} of deep neural networks, there is no theoretical upperbound of $n$ under our \emph{HINNPerf} architecture. \emph{Theoretically}, \emph{HINNPerf} can predict the performance of complex systems with arbitrary number of configuration options. The complete theoretical proof is beyond the scope of this paper, and here we intuitively justify the statement using the main theorem from~\cite{approx_03}:
\begin{theorem}[main theorem]
    Under certain assumptions, it holds that for any function $f: \mathbb{R}^n \to \mathbb{R}$ with given approximation error $\epsilon$, there exists a positive integer $z$, and a fully connected and feed-forward deep neural network of $l = \lceil \log_2 z \rceil$ hidden layers, with $n$ inputs and a single output and with ReLU activation such that:
    \begin{equation}
    z \le \frac{Cn}{\epsilon^2},
    \end{equation}
    where the constant $C$ depends on certain assumptions, but not on $n$.
\end{theorem}
All the assumptions and the complete proof of the above theorem can be found in~\cite{approx_03}. Note that as $n \to \infty$ or $\epsilon \to 0$, we have $z < \infty$ and thus $l < \infty$, implying that for any value of $n$, there always exists a deep FNN that can approximate the function $f$ with proper error $\epsilon$. When it comes to our \emph{HINNPerf} architecture, the $j$-th block employs a deep FNN with $l_j$ hidden layers to approximate the performance subfunction $f_j: \mathbb{R}^{jn} \to \mathbb{R}$ under an error bound $\epsilon \le \sqrt{\frac{Cjn}{2^{l_j}}}$. Therefore, we conclude that in theory \emph{HINNPerf} can approximate the performance function of configurable systems having arbitrary $n$ configuration options without upperbound. And we show that the approximation error of \emph{HINNPerf} is smaller than other baselines in the experiment part of this paper.

\subsection{Model Advantages}
\label{sec:model_analy}

\subsubsection{Beyond the Accuracy}
\label{subsec:beyond_acc}

Although the goal of our method is to improve the prediction accuracy of the performance model, the proposed hierarchical architecture allows us to extract some valuable information about the configurable system. First of all, by measuring the contributions of different partial performance functions $f_j(\mathbf{x}_j)$ to the whole performance function $f(\mathbf{o})$ in Equation~(\ref{equ:perf_decom}), we can obtain some insights about the influences of different interactions on the system performance. If the partial function of lower-order interactions (e.g., $f_1(\mathbf{o})$) has the major contribution, it implies that the interactions among different configuration options are relatively simple and each option tends to affect the system performance independently (i.e., the configuration options are designed to be relatively disentangled). On the contrary, if the partial function of higher-order interactions (e.g., $f_m(\mathbf{x}_m)$) has the major contribution, the feature interactions of the software system might be very complex and more careful configuration decisions should be considered when deploying the configurable system. Second, by computing the significance score of each configuration option in different blocks, we can identify which options have the greatest impact on different partial performance functions $f_j(\mathbf{x}_j)$. Recognizing significant options in different interaction orders (i.e., different blocks) could help users and developers efficiently pinpoint the key performance factors and reveal some potential interaction patterns of the system. The above two advantages make our model more helpful to users and developers than other deep learning methods such as \emph{DeepPerf}~\cite{DeepPerf}. We will show these insights of different real-world subject systems in the experimental part of this paper. 

\subsubsection{Generality}

If we only use one block and empty interaction embedding function (i.e., $m = 1$ and $I_1(\cdot) = \emptyset$), our \emph{HINNPerf} architecture reduces to the \emph{DeepPerf} model~\cite{DeepPerf}. Therefore, \emph{DeepPerf} can be viewed as a special case of our method. From this point of view, \emph{HINNPerf} shows a general deep learning framework for performance prediction of configurable systems. And we believe that more advanced deep learning performance models can be derived from this framework.

\subsection{Training Details}
\label{sec:training_details}

To train our proposed model, we utilize the following technical practice of machine learning:
\begin{itemize}
  \item Dataset separation: As in most machine learning scenarios, we split the performance dataset into training, validation, and testing sets. We use the training set to optimize the model parameters, the validation set to search for the hyperparameters and the testing set to evaluate the model accuracy.
  \item Data normalization: To speed up network training process and facilitate hyperparameter tuning, we normalize the input and the output of the training data. There are two kinds of normalization techniques: maximization and Gaussian normalization. The former normalizes the data in $[0,1]$ and the latter normalizes the data to a Gaussian distribution. We use the validation dataset to choose the better normalization technique. During testing, we normalize the testing dataset by exactly the same parameters of normalization during training, and then denormalize the model predicted output for accuracy evaluation.
  \item Loss function: The loss function of our approach is the mean square error between the real performance values and the predicted output, which is the common regression loss function in machine learning.
  \item Optimization: We utilize the Adam~\cite{adam} algorithm to train our neural network. During the training process, the batch size is equal to the whole sample size of the training dataset since the size of the training data is small for performance prediction problem.
  \item Hyperparameter setting: We conduct a grid search over all hyperparameters of our method. We fix the initial learning rate to be $0.001$ and employ a learning rate schedule that drops the rate by a factor of $10^{-3}$ after every training epoch. We vary the number of blocks in $\{2,3,4,5\}$ (i.e., the hyperparameter $m$ in Equation~(\ref{equ:perf_decom})) since the interactions among different options are at most five-way~\cite{SPLConquer_02,SPLConquer_03,HierarchicalInter}. Also, we choose the number of hidden layers in each block from $\{2,3,4\}$ and fix the number of neurons in each hidden layer to be $128$. The $L_1$ regularization hyperparameter $\lambda$ is chosen from $\{0.001, 0.01, 0.1, 1, 10\}$. The optimal hyperparameter setting is the setting that achieves the smallest validation error. 
\end{itemize}

Finally, we implement our proposed method \emph{HINNPerf} using Python 3.6 and Tensorflow 1.13.0~\cite{Tensorflow} under a machine with one NVIDIA GeForce GTX 1080 GPU and one Intel(R) Core(TM) i7-6800K CPU 3.40GHz.

\section{Evaluation}

The core of this section is to evaluate whether our method can better model the complex interactions in real-world configurable systems and accurately predict the performance with a small training sample (i.e., the two challenges mentioned in Section~\ref{sec:Intro}). Specifically, we aim at answering the following research questions (RQ):
\begin{itemize}
  \item \textbf{RQ1}: How accurate is our approach in predicting performance of configurable software systems?
  \item \textbf{RQ2}: Are the proposed hierarchical interaction embedding architecture and the hierarchical $L_1$ regularization strategy necessary for improving prediction accuracy?
  \item \textbf{RQ3}: How do the $L_1$ regularization hyperparameter $\lambda$ and the interaction order $m$ affect the efficacy of \emph{HINNPerf}?
  \item \textbf{RQ4}: In addition to predicting performance values, can \emph{HINNPerf} provide some insights on configuration coupling and help users and developers find out significant configuration options in the performance model?
  \item \textbf{RQ5}: What is the time cost of training and testing the \emph{HINNPerf} model for performance prediction?
\end{itemize}

We next conduct different experiments to answer the RQs above, comprehensively demonstrating the effectiveness, robustness and practicality of our approach.

\subsection{Evaluation Metric}

As described in Section~\ref{sec:training_details}, we use the training and validation datasets to generate a performance model for each system, and then use this model to predict the performance values of configurations on the testing dataset. Similar to~\cite{SPLConquer_03,DECART,PerLasso,DeepPerf}, we utilize the mean relative error (MRE) to evaluate the model's prediction accuracy:
\begin{equation}
MRE = \frac{1}{|\mathcal{T}|} \sum_{\mathbf{o} \in \mathcal{T}} \frac{|f(\mathbf{o}) - \hat{f}(\mathbf{o})|}{f(\mathbf{o})} \times 100,
\end{equation}
where $\mathcal{T}$ denotes the testing dataset, $f(\mathbf{o})$ is the actual performance value of configuration $\mathbf{o}$, and $\hat{f}(\mathbf{o})$ is the predicted performance value of configuration $\mathbf{o}$.

\subsection{Subject Systems}

\begin{table}[htbp]
  \caption{Overview of the subject systems}
  \label{tab:subject_sys}
  \resizebox{\textwidth}{!}{
  \begin{tabular}{|l|l|c|c|c|l|c|c|}
    \cline{1-8}
    System    &  Domain                   &  $|\mathcal{B}|$  &  $|\mathcal{N}|$  &  $|\mathcal{C}|$  &  Performance Value  &  Range              &  Variance    \\
    \cline{1-8}
    x264      &  Video Encoder            &  16               &  0                &  1152             &  Encoding time      &  [244, 822]         &  $2.64 \times 10^{4}$    \\
    BDB-J     &  Database System          &  26               &  0                &  180              &  Response time      &  [2960, 16531]      &  $1.76 \times 10^{7}$    \\
    LRZIP     &  File Archive Utility     &  19               &  0                &  432              &  Compression time   &  [37470, 5811280]   &  $1.35 \times 10^{12}$   \\
    VP9       &  Video Encoder            &  42               &  0                &  216000           &  Encoding time      &  [41, 100]          &  $5.12 \times 10^{2}$    \\
    POLLY     &  Code Optimizer           &  40               &  0                &  60000            &  Runtime            &  [4, 32]            &  $4.16 \times 10^{1}$                 \\
    Dune MGS  &  Multi-Grid Solver        &  8                &  3                &  2304             &  Solving time       &  [4422, 58092]      &  $1.35 \times 10^{7}$    \\
    $\text{HIPA}^{cc}$ & Image Processing &  31               &  2                &  13485            &  Solving time       &  [21, 122]          &  $2.41 \times 10^{2}$    \\
    HSMGP     &  Stencil-Grid Solver      &  11               &  3                &  3456             &  Solving time       &  [100, 4397]        &  $6.78 \times 10^{5}$    \\
    JavaGC    &  Garbage Collector        &  12               &  23               &  166975           &  Collection time    &  [370, 80004]       &  $5.84 \times 10^{6}$    \\
    SaC       &  Compiler                 &  53               &  7                &  62523            &  Compilation time   &  [1, 492]           &  $3.25 \times 10^{2}$    \\
    \cline{1-8}
  \end{tabular}}
  {\raggedright \footnotesize{} \par}
  {\raggedright \small{$|\mathcal{B}|$: number of binary options; $|\mathcal{N}|$: number of numeric options; $|\mathcal{C}|$: total number of valid configurations measured for each system; Range \& Variance: the range and the variance of performance values in each dataset.} \par}
\end{table}

In our experiments, we consider 10 real-world configurable software systems from different domains, including video encoders, database systems, multi-grid solvers, image processing frameworks, compilers, etc. Five of these systems only have binary configuration options and the other five systems have both binary and numeric configuration options. Different systems have different sizes (from 45 thousands to more than 300 thousands lines of code) and are written in different programming languages (Java, C, and C++). Table~\ref{tab:subject_sys} provides an overview of the subject systems. According to previous work~\cite{SPLConquer_02,SPLConquer_03,DeepPerf,distance-sampling}, the performance datasets of the ten subject systems are collected in the following way:
\begin{itemize}
    \item x264 is a video encoder for the H.264 compression format. Relevant configuration options included the number of reference frames, enabling or disabling the default entropy encoder, and the number of frames for ratecontrol and lookahead. We have measured the time to encode the Sintel trailer (734 MB) on an Intel Core Q6600 with 4 GB RAM (Ubuntu 14.04).
    \item BDB-J is the Java version of the Berkeley Database. we use Oracle's standard benchmark to measure the performance of BDB-J. The workload produced by the benchmarks is a typical sequence of database operations.
    \item LRZIP is a file compression tool. We consider configuration options that define, for instance, the compression level and the use of encryption. We used the uiq28 generator to generate a file (632 MB), and we measured the time for compressing this file with version 0.600 on a machine with AMD Athlon64 Dual Core, 2 GB RAM (Debian GNU/Linux 6).
    \item VPXENC (VP9) is a video encoder that uses the VP9 video coding format. It offers different configuration options, such as adjusting the quality, the bitrate of the coded video, and the number of threads to use. We measured the encoding time of 2 seconds from the Big Buck Bunny trailer on an Intel Xeon E5-2690 and 64 GB RAM (Ubuntu 16.04).
    \item POLLY is a loop optimizer that rests on top of LLVM. POLLY provides various configuration options that define, for example, whether code should be parallelized or the choice of the tile size. We used POLLY version 3.9, LLVM version 4.0.0, and Clang version 4.0.0. As benchmark, we used the gemm program from polybench and measured its runtime on an Intel Xeon E5-2690 and 64 GB RAM (Ubuntu 16.04).
    \item Dune MGS is a geometric multi-grid solver based on the Dune framework. The framework provides algorithms for smoothing and solving Poisson equations on structured grids. Binary options include several smoother and solver algorithms. Numeric options include different grid sizes and pre- and post-smoothing steps. We measured the time to solve Poisson's equation on a Dell OptiPlex-9020 with an Intel i5-4570 Quad Code and 32 GB RAM (Ubuntu 13.4).
    \item $\text{HIPA}^{cc}$ is an image processing acceleration framework, which generates efficient low-level code from a high-level specification. Binary options are, among others, the kind of memory to be used (e.g., texture vs. local). The number of pixels calculated per thread is an example of a numeric option. We measured the time needed for solving a test set of partial differential equations on an nVidia Tesla K20 card with 5GB RAM and 2496 cores (Ubuntu 14.04).
    \item HSMGP is a highly scalable multi-grid solver for large-scale data sets. Binary options include in-place conjugate gradient and in-place algebraic multi-grid solvers. Numeric options include the number of smoothing steps and the number of nodes used for computing the solution. As a benchmark, we performed a multi-grid iteration of solving Poisson's equation. We executed the benchmark runs on JuQueen, a Blue Gene/Q system, located at the Julich Supercomputing Center, Germany.
    \item JavaGC is the Java garbage collector (version 7) with several options for adaptive garbage-collection boundary and size policies. For measurement, we executed the DaCapo benchmark suite on a computing cluster consisting of 16 nodes each equipped with an Intel Xeon E5-2690 Ivy Bridge having 10 cores and 64 GB RAM (Ubuntu 14.04).
    \item SaC is a variant of C for high-performance computing based on stateless arrays. The SaC compiler implements a large number of high-level and low-level optimizations to tune high-level programs for efficient parallel executions. The compiler is highly configurable, allowing users to select various optimizations and to customize the optimization effort (e.g., optimization cycles and loop-unrolling threshold). As benchmark, we compile and execute an n-body simulation shipped with the compiler, measuring the execution time of the simulation at different optimization levels. We executed all benchmarks on an 8 core Intel i7-2720QM machine with 8 GB RAM (Ubuntu 12.04).
\end{itemize}

\subsection{RQ1: Performance Prediction Comparison}\label{sec: results}

\subsubsection{Baselines}

As introduced in Section~\ref{sec:Intro}, many learning methods have been proposed to predict performance values of configurable systems with binary and/or numeric options, including \emph{SPLConqueror}~\cite{SPLConquer_02,SPLConquer_03}, \emph{Fourier} methods~\cite{Fourier,PerLasso}, \emph{DECART}~\cite{DECART}, and \emph{DeepPerf}~\cite{DeepPerf}. Besides, traditional machine learning methods such as Random Forests~\cite{RDF} (\emph{RF}) can also be used for the performance prediction task. For configurable systems with only binary options, \emph{DECART} and \emph{DeepPerf} are recent advanced methods and can achieve higher prediction accuracy than others~\cite{DECART,DeepPerf}. Therefore, to save experiment time, we only compare the proposed method with three baselines \emph{DECART}, \emph{DeepPerf} and \emph{RF} on binary systems. For configurable systems with both binary and numeric options, we compare the effectiveness of our \emph{HINNPerf} model with that of \emph{SPLConqueror}, \emph{DeepPerf} and \emph{RF}, since the \emph{DECART} model cannot deal with numeric options.

\subsubsection{Setup}

Following the experiment setup in~\cite{DeepPerf}, we \emph{randomly} select a certain number of configurations and their corresponding performance values to construct the training and validation datasets (sample\footnote{We use 67\% of sample for training and 33\% for validation.}), and use the remaining configuration measurements as the testing dataset. For each subject system, we test the model accuracy under different sample sizes. The sample sizes of each binary system are $n, 2n, 4n, 6n$, where $n$ is the number of binary options of each system (shown in the column $|\mathcal{B}|$ of Table~\ref{tab:subject_sys}); for each binary-numeric system, we choose the same sample sizes that \emph{SPLConqueror} suggested~\cite{SPLConquer_03}. Same as in~\cite{DeepPerf}, for \emph{DECART}, \emph{DeepPerf}, \emph{RF} and our \emph{HINNPerf} methods, we repeat the random sampling, training and testing process 30 times, and then report the mean and the $95\%$ confidence interval of each method's MRE obtained after 30 experiments. For \emph{SPLConqueror}, however, we cannot choose a random sample since it combines different sampling heuristics for binary options (e.g., option-wise (OW), pair-wise (PW), etc), and several experimental design methodologies (e.g., Plackett-Burman (PBD), Random Design (RD), etc) for numeric options to predict system performance. Each combination of sampling heuristics and non-random experimental designs yields a unique sample, implying that repeating 30 experiments for \emph{SPLConqueror} produces exactly the same results and thus the variance of the 30 experimental results is zero. Hence, we only report the MRE of \emph{SPLConqueror} on the testing dataset since the $95\%$ confidence interval is also zero when the variance is zero~\cite{DeepPerf}.
Moreover, we check whether there is a statistically significant difference between the results of the best and the second best methods by using the Wilcoxon rank sum test~\cite{ranksum}, a non-parametric test that compares different distributions. When the $p$-value is smaller than $0.05$, we consider that there is a significant difference and report the significantly better method.

To replicate the results of \emph{SPLConqueror}, \emph{DECART} and \emph{DeepPerf}, we utilize the code published on their supplement websites and run their models with the best hyperparameter settings~\cite{SPLConquer_03,DECART,DeepPerf}. For the \emph{RF} method, we use the function \emph{RandomForestRegressor} in scikit-learn package~\cite{sklearn} to perform model training and prediction. To select the best hyperparameters for \emph{RF}, we construct the grid search by varying four hyperparameters in \emph{RandomForestRegressor}: 10 values of {\tt n\_estimators} ranging from 10 to 1000, 10 values of {\tt min\_samples\_leaf} ranging from 0.05 to 0.5, 10 values of {\tt max\_leaf\_nodes} ranging from 2 to 100 and the {\tt max\_features} are {\tt auto}, {\tt sqrt} or {\tt log2}. With this setting, the hyperparameter space contains 3000 different combinations of hyperparameter values. Hence, we believe it is sufficient enough to find the optimal setting for \emph{RF}.

\begin{table*}
  \centering
    \caption{Performance prediction results of ten subject systems}
    \label{tab:result}
    \resizebox{\textwidth}{!}{
    \begin{tabular}{|l|l|c|c|c|c|c|c|c|c|c|c|c|}
      \cline{1-13}
      Subject                                 &  Sample             &  \multicolumn{2}{c|}{\emph{SPLConqueror}}            &  \multicolumn{2}{c|}{\emph{DECART}}  &  \multicolumn{2}{c|}{\emph{DeepPerf}}  &  \multicolumn{2}{c|}{\emph{RF}}        &  \multicolumn{2}{c|}{\emph{HINNPerf}}  &  Better             \\
      \cline{3-12}
      System                                  &  Size ($\%$)        &  Sampling Heuristic                &  Mean           &  Mean                     &  Margin  &  Mean                       &  Margin  &  Mean                       &  Margin  &  Mean                       &  Margin  &  Method             \\
      \cline{1-13}
      \multirow{4}{*}{x264}                   &  $n~(1.39\%)$       &  --                                &  --             &  17.71                    &  3.87    &  \underline{10.43}          &  2.28    &  18.19                      &  2.24    &  \textbf{9.68}              &  1.40    &  \emph{HINNPerf}    \\                       
                                              &  $2n~(2.78\%)$      &  --                                &  --             &  9.31                     &  1.30    &  \underline{3.61}           &  0.54    &  12.39                      &  0.99    &  \textbf{3.00}              &  0.31    &  \emph{HINNPerf}    \\
                                              &  $4n~(5.56\%)$      &  --                                &  --             &  4.26                     &  0.47    &  \underline{1.49}           &  0.38    &  8.47                       &  0.46    &  \textbf{0.98}              &  0.16    &  \emph{HINNPerf}    \\
                                              &  $6n~(8.34\%)$      &  --                                &  --             &  2.05                     &  0.25    &  \underline{0.92}           &  0.15    &  7.35                       &  0.32    &  \textbf{0.42}              &  0.06    &  \emph{HINNPerf}    \\
      \cline{1-13}
      \multirow{4}{*}{BDB-J}                  &  $n~(14.44\%)$      &  --                                &  --             &  10.04                    &  4.67    &  \underline{7.25}           &  4.21    &  16.64                      &  2.35    &  \textbf{4.94}              &  2.27    &  Same               \\                       
                                              &  $2n~(28.88\%)$     &  --                                &  --             &  2.23                     &  0.16    &  \underline{2.07}           &  0.32    &  9.17                       &  1.23    &  \textbf{2.06}              &  0.15    &  Same               \\
                                              &  $4n~(57.76\%)$     &  --                                &  --             &  1.72                     &  0.09    &  \underline{1.67}           &  0.12    &  5.04                       &  0.54    &  \textbf{1.53}              &  0.07    &  Same               \\
                                              &  $6n~(86.64\%)$     &  --                                &  --             &  1.57                     &  0.13    &  \underline{1.49}           &  0.13    &  3.85                       &  0.39    &  \textbf{1.42}              &  0.14    &  Same               \\
      \cline{1-13}
      \multirow{4}{*}{LRZIP}                  &  $n~(4.40\%)$       &  --                                &  --             &  \underline{198.79}       &  92.61   &  \textbf{173.11}            &  50.32   &  229.59                     &  57.47   &  214.52                     &  84.82   &  Same               \\                       
                                              &  $2n~(8.80\%)$      &  --                                &  --             &  \underline{56.73}        &  20.36   &  67.23                      &  27.04   &  151.58                     &  32.53   &  \textbf{48.44}             &  8.82    &  Same               \\
                                              &  $4n~(17.60\%)$     &  --                                &  --             &  \underline{19.06}        &  2.98    &  19.76                      &  2.61    &  96.44                      &  19.86   &  \textbf{16.53}             &  2.90    &  Same               \\
                                              &  $6n~(26.40\%)$     &  --                                &  --             &  \underline{11.22}        &  2.07    &  12.75                      &  2.54    &  69.94                      &  17.01   &  \textbf{10.05}             &  2.03    &  Same               \\
      \cline{1-13}
      \multirow{4}{*}{VP9}                    &  $n~(0.02\%)$       &  --                                &  --             &  3.15                     &  0.13    &  \underline{3.12}           &  0.12    &  14.54                      &  2.86    &  \textbf{3.09}              &  0.21    &  Same               \\                       
                                              &  $2n~(0.04\%)$      &  --                                &  --             &  \underline{2.23}         &  0.23    &  2.38                       &  0.13    &  15.54                      &  1.36    &  \textbf{1.85}              &  0.15    &  \emph{HINNPerf}    \\
                                              &  $4n~(0.08\%)$      &  --                                &  --             &  \underline{1.21}         &  0.14    &  1.83                       &  0.14    &  14.19                      &  0.42    &  \textbf{0.73}              &  0.08    &  \emph{HINNPerf}    \\
                                              &  $6n~(0.12\%)$      &  --                                &  --             &  \underline{0.86}         &  0.09    &  1.06                       &  0.15    &  13.38                      &  0.42    &  \textbf{0.44}              &  0.02    &  \emph{HINNPerf}    \\
      \cline{1-13}
      \multirow{4}{*}{POLLY}                  &  $n~(0.07\%)$       &  --                                &  --             &  17.98                    &  3.79    &  \textbf{15.61}             &  2.68    &  24.38                      &  1.34    &  \underline{16.33}          &  2.52    &  Same               \\                       
                                              &  $2n~(0.14\%)$      &  --                                &  --             &  \underline{5.84}         &  1.11    &  \textbf{5.36}              &  0.85    &  18.04                      &  1.62    &  6.20                       &  0.89    &  Same               \\
                                              &  $4n~(0.28\%)$      &  --                                &  --             &  \underline{4.20}         &  0.62    &  4.82                       &  0.68    &  14.51                      &  1.82    &  \textbf{4.14}              &  0.48    &  Same               \\
                                              &  $6n~(0.42\%)$      &  --                                &  --             &  \textbf{3.03}            &  0.52    &  3.60                       &  0.41    &  13.88                      &  1.14    &  \underline{3.50}           &  0.22    &  \emph{DECART}      \\
      \cline{1-13}
      \multirow{4}{*}{Dune MGS}               &  $49~(2.13\%)$      &  OW RD                             &  20.1           &  --                       &  --      &  \underline{15.73}          &  0.90    &  17.73                      &  0.64    &  \textbf{13.43}             &  0.84    &  \emph{HINNPerf}    \\                       
                                              &  $78~(3.39\%)$      &  PW RD                             &  22.1           &  --                       &  --      &  \underline{13.67}          &  0.82    &  18.19                      &  0.47    &  \textbf{11.93}             &  0.67    &  \emph{HINNPerf}    \\
                                              &  $384~(16.67\%)$    &  PW PBD(49,7)                      &  11             &  --                       &  --      &  \underline{7.20}           &  0.18    &  8.29                       &  0.11    &  \textbf{6.74}              &  0.19    &  \emph{HINNPerf}    \\
                                              &  $600~(26.04\%)$    &  PW PBD(125,5)                     &  8.3            &  --                       &  --      &  \underline{6.44}           &  0.20    &  7.46                       &  0.09    &  \textbf{5.86}              &  0.13    &  \emph{HINNPerf}    \\
      \cline{1-13}
      \multirow{4}{*}{$\text{HIPA}^{cc}$}     &  $261~(1.94\%)$     &  OW RD                             &  14.2           &  --                       &  --      &  \underline{9.39}           &  0.37    &  14.49                      &  0.27    &  \textbf{7.24}              &  0.36    &  \emph{HINNPerf}    \\                       
                                              &  $528~(3.92\%)$     &  OW PBD(125,5)                     &  13.8           &  --                       &  --      &  \underline{6.38}           &  0.44    &  11.38                      &  0.19    &  \textbf{4.55}              &  0.20    &  \emph{HINNPerf}    \\
                                              &  $736~(5.46\%)$     &  OW PBD(49,7)                      &  13.9           &  --                       &  --      &  \underline{5.06}           &  0.35    &  9.87                       &  0.14    &  \textbf{3.59}              &  0.12    &  \emph{HINNPerf}    \\
                                              &  $1281~(9.50\%)$    &  PW RD                             &  13.9           &  --                       &  --      &  \underline{3.75}           &  0.26    &  7.54                       &  0.10    &  \textbf{2.81}              &  0.06    &  \emph{HINNPerf}    \\
      \cline{1-13}
      \multirow{4}{*}{HSMGP}                  &  $77~(2.23\%)$      &  OW RD                             &  \textbf{4.5}   &  --                       &  --      &  6.76                       &  0.87    &  34.48                      &  3.17    &  \underline{5.59}           &  0.75    &  Same    \\                       
                                              &  $173~(5.01\%)$     &  PW RD                             &  \textbf{2.8}   &  --                       &  --      &  3.60                       &  0.2     &  22.31                      &  1.16    &  \underline{3.02}           &  0.18    &  \emph{SPLConqueror}    \\
                                              &  $384~(11.11\%)$    &  OW PBD(49,7)                      &  \underline{2.2}&  --                       &  --      &  2.53                       &  0.13    &  14.78                      &  0.37    &  \textbf{1.98}              &  0.15    &  \emph{HINNPerf}    \\
                                              &  $480~(13.89\%)$    &  OW PBD(125,5)                     &  \textbf{1.7}   &  --                       &  --      &  2.24                       &  0.11    &  12.56                      &  0.68    &  \underline{1.75}           &  0.08    &  Same    \\
      \cline{1-13}
      \multirow{4}{*}{JavaGC}                 &  $855~(0.51\%)$     &  OW PBD(125,5)                     &  21.9           &  --                       &  --      &  \underline{21.83}          &  7.07    &  50.59                      &  2.27    &  \textbf{15.99}             &  1.17    &  \emph{HINNPerf}    \\                       
                                              &  $2571~(1.54\%)$    &  PW PBD(49,7)                      &  28.2           &  --                       &  --      &  \underline{16.48}          &  6.59    &  33.40                      &  0.71    &  \textbf{10.02}             &  0.44    &  \emph{HINNPerf}    \\
                                              &  $3032~(1.82\%)$    &  PW RD                             &  24.6           &  --                       &  --      &  \underline{12.76}          &  0.56    &  31.29                      &  0.66    &  \textbf{9.74}              &  0.54    &  \emph{HINNPerf}    \\
                                              &  $5312~(3.18\%)$    &  PW PBD(125,5)                     &  18.8           &  --                       &  --      &  \underline{11.03}          &  0.45    &  23.90                      &  0.47    &  \textbf{7.12}              &  0.24    &  \emph{HINNPerf}    \\
      \cline{1-13}
      \multirow{4}{*}{SaC}                    &  $2060~(3.29\%)$    &  OW RD                             &  21.1           &  --                       &  --      &  \underline{15.83}          &  1.25    &  62.72                      &  3.02    &  \textbf{13.50}             &  0.94    &  \emph{HINNPerf}    \\                       
                                              &  $2295~(3.67\%)$    &  OW PBD(125,5)                     &  20.3           &  --                       &  --      &  \underline{19.25}          &  6.03    &  59.37                      &  1.61    &  \textbf{12.94}             &  0.79    &  \emph{HINNPerf}    \\
                                              &  $2499~(4.00\%)$    &  OW PBD(49,7)                      &  \underline{16} &  --                       &  --      &  16.73                      &  1.13    &  58.07                      &  1.90    &  \textbf{12.37}             &  0.70    &  \emph{HINNPerf}      \\
                                              &  $3261~(5.22\%)$    &  PW RD                             &  30.7           &  --                       &  --      &  \underline{15.64}          &  1.18    &  51.39                      &  2.02    &  \textbf{11.48}             &  0.73    &  \emph{HINNPerf}    \\
      \cline{1-13}
    \end{tabular}}
    {\raggedright \footnotesize{} \par}
    {\raggedright \small{The best result is in \textbf{bold} and the second best result is \underline{underlined}.} \par}
    {\raggedright \small{Mean: mean of the MREs seen in 30 experiments. Margin: margin of the $95\%$ confidence interval of the MREs in 30 experiments.} \par}
    {\raggedright \small{Better Method is chosen between the \textbf{best method} and the \underline{second best method} using Wilcoxon test on MREs of 30 experiments with $5\%$ significance level.} \par}
    {\raggedright \small{The percentages ($\%$) after each sample size show the ratio of the training sample size to the total number of data samples measured for each system.} \par}
\end{table*}

\subsubsection{Results}

Table~\ref{tab:result} summarizes the performance prediction results of each approach on the 10 subject systems with small and large sample sizes, from which we obtain the following observations:
\begin{itemize}
  \item Our method clearly outperforms other approaches on most of the subject systems, showing the superiority of hierarchical interaction network design for performance prediction. Specifically, \emph{HINNPerf} achieves best prediction accuracy on 7 out of 10 systems, reducing the MRE by an average of \textbf{22.67\%} compared to the second best results of other baselines. Remarkably, \emph{HINNPerf} achieves the best \textbf{54.35\%} improvement on the x264 system with $6n$ sample size. Among the 7 winning systems, the Wilcoxon test shows that \emph{HINNPerf} significantly outperforms the second best method on 5 of them. For the other 3 systems LRZIP, POLLY, and HSMGP, \emph{HINNPerf} and other methods perform quite similarly without significant difference between them. On the other hand, we observe that the sample size of $n$ in systems with only binary options seems too small to train each performance model for achieving good accuracy. As we can see from Table~\ref{tab:result}, the prediction MREs of all models under sample size $n$ are more than twice higher than the MREs under sample size $2n$ on the x264, BDB-J, LRZIP, and POLLY systems, respectively. What’s worse, the MREs of all models under the sample size $n$ of the LRZIP system are more than $100\%$, implying that all performance models cannot work when trained in this case. Therefore, the prediction results under sample size $n$ may not really reveal the efficacy of each performance model.
  \item We conduct the Wilcoxon test on the prediction results of our method and the most related deep learning method \emph{DeepPerf}~\cite{DeepPerf}. It shows that \emph{HINNPerf} significantly outperforms \emph{DeepPerf} over all of the 5 systems with both binary and numeric configuration options (i.e., Dune MGS, $\text{HIPA}^{cc}$, HSMGP, JavaGC, and SaC). Since binary-numeric interactions are usually more complicated than binary-binary interactions~\cite{SPLConquer_03,DeepPerf}, the result of the Wilcoxon test demonstrates that our method can better model the influences of complex interactions on system performance.
  \item Since the major contribution of the performance model is to accurately predict the performance when only a small subset of data samples is avaliable for training, it is important to evaluate the accuracy of each performance model under a small sample size. From Table~\ref{tab:result}, except for the BDB-J system, the smallest sample size used for training is less than $5\%$ of the total number of data samples measured for each system. Using these small data subsets, our \emph{HINNPerf} model achieves a prediction accuracy of more than $83\%$ (i.e., MRE less than $17\%$) on 9 out of 10 systems, with accuracy of 5 systems (i.e., x264, BDB-J, VP9, $\text{HIPA}^{cc}$, and HSMGP) more than $90\%$. Besides, our method outperforms other baselines on 7 out of 10 systems under the smallest sample size. Therefore, we conclude that our proposed \emph{HINNPerf} model can still attain a good accuracy when trained with only a small subset of data samples. Furthermore, to obtain the same level of accuracy, \emph{HINNPerf} needs fewer data than other methods. For example, with the system JavaGC, \emph{HINNPerf} only needs 2571 sample to achieve a prediction MRE of $10.02$\% (i.e., accuracy of $89.98$\%) whilst \emph{DeepPerf} needs 5312 (more than 2 times) sample to obtain the similar prediction accuracy. These observations demonstrate the effectiveness of our approach in small sample predictions.
  \item We observe that there are some differences in the prediction MREs of different systems. For example, the prediction MRE of all performance models on the LRZIP system is much higher than other subject systems, probably due to the large scales and variance of the performance value of the LRZIP system (see Table~\ref{tab:subject_sys}). It seems that the scales and variance of the performance metrics may be an issue affecting the prediction accuracy of the performance model, and the data with larger variance is more difficult to predict. This observation inspires us that, in practice, it is better to properly control the data variance when measuring the performance of a configurable system to train the performance model. Since the goal of this paper is to design a novel performance model that outperforms other state-of-the-art models, we demonstrate the effectiveness of our model only using the public benchmark datasets~\cite{SPLConquer_02,SPLConquer_03,DeepPerf,distance-sampling} without changing the scales and variance of the datasets. Instead, we show that our \emph{HINNPerf} model achieves better accuracy on most systems with performance metrics having different scales and variance.
\end{itemize}

In summary, compared to existing methods, our proposed architecture \emph{HINNPerf} provides a more effective and accurate deep learning performance model to address the interaction modeling and the small sample challenges mentioned in Section~\ref{sec:Intro}.

\subsection{RQ2: Effectiveness of Architecture Design}

Our method consists of two key components, the hierarchical interaction embedding (Equation~(\ref{equ:perf_decom})) and the hierarchical $L_1$ regularization, as shown in Figure~\ref{fig:block}. In this experiment, we aim to evaluate whether these two components are necessary to improve performance prediction accuracy, or they can be substituted by other techniques. To show the effectiveness of each component, we compare \emph{HINNPerf} with following design alternatives:
\begin{itemize}
  \item \emph{MB-FNN}: We remove the hierarchical interaction embedding from the \emph{HINNPerf} architecture. In this way, the neural network reduces to a multi-block FNN with $L_1$ regularization on the first layer of each block.
  \item \emph{L2-HINN}: We replace the $L_1$ regularization of \emph{HINNPerf} with $L_2$ regularization.
  \item \emph{Plain-HINN}: This architecture is simply the \emph{HINNPerf} model without regularization.
  \item \emph{Dropout-HINN}: We replace the $L_1$ regularization of \emph{HINNPerf} with the dropout technique~\cite{dropout}.
\end{itemize}
We evaluate each architecture on the 10 subject systems and compare the prediction results with the \emph{HINNPerf} model.

\subsubsection{Setup} 

For all the alternative models, we also use the same grid search strategy to tune the hyperparameters. All the hyperparameter settings are the same as those for \emph{HINNPerf}. Except that, for \emph{Dropout-HINN}, the search range for the dropout rate is $\{0.1, 0.25, 0.5, 0.75, 0.9\}$ since the dropout rate needs to be $\le 1$.

\subsubsection{Results}

Table~\ref{tab:ablation} shows the prediction MREs of \emph{HINNPerf} and 4 alternative methods. 
Overall, \emph{HINNPerf} achieves the best accuracy on most subject systems. Comparing the results of \emph{HINNPerf} and \emph{MB-FNN}, we find that removing the interaction embedding from the architecture causes a large accuracy drop overall (e.g., on average, 58\% accuracy drop on VP9 and 68\% accuracy drop on POLLY), which demonstrates that the hierarchical interaction embedding plays an important role in improving the prediction accuracy of our performance model. Hence, we believe that it is worth introducing the embedding method to further improve deep learning performance models, confirming our discussion in Section~\ref{subsec:inter_embed}. 

On the other hand, removing the regularization technique from \emph{HINNPerf} (i.e., the \emph{Plain-HINN} model) causes the most significant accuracy drops, since a deep neural network is prone to overfit on training data. The \emph{L2-HINN} and \emph{Dropout-HINN} methods perform much better than \emph{Plain-HINN} but are still less effective than \emph{HINNPerf}. These observations show that the hierarchical $L_1$ regularization architecture may be the most important component for our approach.

In conclusion, both the hierarchical interaction embedding and the hierarchical $L_1$ regularization techniques are necessary for prediction accuracy improvement. The full \emph{HINNPerf} architecture is the most accurate and effective peformance model for configurable systems when compared to other alternatives.

\begin{table*}
\centering
  \caption{Comparison between \emph{HINNPerf} and other alternatives}
  \label{tab:ablation}
  \resizebox{0.75\textwidth}{!}{
  \begin{tabular}{|l|l|c|c|c|c|c|c|c|c|c|c|}
    \cline{1-12}
    Subject                                 &  Sample  &  \multicolumn{2}{c|}{\emph{HINNPerf}}           &  \multicolumn{2}{c|}{\emph{MB-FNN}}  &  \multicolumn{2}{c|}{\emph{L2-HINN}}  &  \multicolumn{2}{c|}{\emph{Plain-HINN}}  &  \multicolumn{2}{c|}{\emph{Dropout-HINN}} \\
    \cline{3-12}
    System                                  &  Size    &  Mean                       &  Margin           &  Mean                     &  Margin  &  Mean                      &  Margin  &  Mean                         &  Margin  &  Mean                         &  Margin  \\
    \cline{1-12}
    \multirow{3}{*}{x264}                   &  $2n$    &  \textbf{3.00}              &  0.31             &  3.09                     &  0.28    &  4.55                      &  0.48    &  9.09                         &  0.98    &  7.61                         &  0.54    \\ 
                                            &  $4n$    &  \textbf{0.98}              &  0.16             &  1.07                     &  0.16    &  1.90                      &  0.17    &  4.49                         &  0.21    &  4.47                         &  0.34    \\ 
                                            &  $6n$    &  \textbf{0.42}              &  0.06             &  0.54                     &  0.08    &  1.02                      &  0.09    &  3.18                         &  0.16    &  3.42                         &  0.16    \\ 
    \cline{1-12}
    \multirow{3}{*}{BDB-J}                  &  $2n$    &  \textbf{2.06}              &  0.15             &  2.40                     &  0.30    &  2.26                      &  0.23    &  5.70                         &  0.84    &  4.27                         &  0.48    \\ 
                                            &  $4n$    &  \textbf{1.53}              &  0.07             &  \textbf{1.53}            &  0.08    &  1.61                      &  0.09    &  2.56                         &  0.25    &  2.77                         &  0.15    \\ 
                                            &  $6n$    &  1.42                       &  0.14             &  \textbf{1.30}            &  0.16    &  1.68                      &  0.18    &  1.86                         &  0.21    &  3.09                         &  0.25    \\     
    \cline{1-12}
    \multirow{3}{*}{LRZIP}                  &  $2n$    &  \textbf{48.44}             &  8.82             &  58.21                    &  15.91   &  83.85                     &  19.82   &  82.97                        &  13.48   &  77.08                        &  17.50   \\ 
                                            &  $4n$    &  \textbf{16.53}             &  2.90             &  22.01                    &  6.29    &  26.31                     &  6.38    &  47.95                        &  12.55   &  42.10                        &  10.55   \\ 
                                            &  $6n$    &  \textbf{10.05}             &  2.03             &  12.07                    &  4.75    &  10.32                     &  2.11    &  28.03                        &  9.83    &  18.95                        &  4.73    \\ 
    \cline{1-12}
    \multirow{3}{*}{VP9}                    &  $2n$    &  \textbf{1.85}              &  0.15             &  2.21                     &  0.24    &  2.40                      &  0.15    &  6.13                         &  0.47    &  4.17                         &  0.23    \\ 
                                            &  $4n$    &  \textbf{0.73}              &  0.08             &  1.22                     &  0.10    &  1.32                      &  0.06    &  3.63                         &  0.12    &  2.88                         &  0.12    \\ 
                                            &  $6n$    &  \textbf{0.44}              &  0.02             &  0.83                     &  0.08    &  0.90                      &  0.06    &  3.08                         &  0.11    &  2.35                         &  0.07    \\ 
    \cline{1-12}
    \multirow{3}{*}{POLLY}                  &  $2n$    &  \textbf{6.20}              &  0.89             &  14.61                    &  3.13    &  14.43                     &  1.49    &  22.71                        &  2.32    &  20.51                        &  1.73    \\ 
                                            &  $4n$    &  \textbf{4.14}              &  0.48             &  6.31                     &  1.66    &  5.66                      &  0.66    &  14.96                        &  2.07    &  11.72                        &  1.40    \\ 
                                            &  $6n$    &  \textbf{3.50}              &  0.22             &  4.15                     &  0.38    &  3.80                      &  0.27    &  10.55                        &  0.78    &  7.86                         &  1.16    \\ 
    \cline{1-12}
    \multirow{4}{*}{Dune MGS}               &  49      &  13.43                      &  0.84             &  \textbf{12.86}           &  2.23    &  13.09                     &  0.82    &  13.72                        &  1.22    &  13.28                        &  0.72    \\ 
                                            &  78      &  11.93                      &  0.67             &  12.03                    &  0.97    &  \textbf{11.83}            &  0.63    &  12.82                        &  1.09    &  12.45                        &  0.53    \\ 
                                            &  384     &  \textbf{6.74}              &  0.19             &  7.19                     &  0.34    &  6.83                      &  0.18    &  7.18                         &  0.24    &  7.56                         &  0.18    \\
                                            &  600     &  \textbf{5.86}              &  0.13             &  6.36                     &  0.39    &  6.08                      &  0.16    &  6.14                         &  0.18    &  6.65                         &  0.10    \\ 
    \cline{1-12}
    \multirow{4}{*}{$\text{HIPA}^{cc}$}     &  261     &  \textbf{7.24}              &  0.36             &  7.28                     &  0.37    &  7.53                      &  0.29    &  10.32                        &  0.94    &  9.19                         &  0.39    \\ 
                                            &  528     &  \textbf{4.55}              &  0.20             &  4.67                     &  0.31    &  4.60                      &  0.19    &  7.30                         &  0.19    &  5.71                         &  0.27    \\ 
                                            &  736     &  \textbf{3.59}              &  0.12             &  3.63                     &  0.16    &  3.66                      &  0.08    &  6.04                         &  0.22    &  4.49                         &  0.16    \\
                                            &  1281    &  \textbf{2.81}              &  0.06             &  2.86                     &  0.07    &  2.82                      &  0.05    &  4.07                         &  0.20    &  3.15                         &  0.10    \\ 
    \cline{1-12}
    \multirow{4}{*}{HSMGP}                  &  77      &  \textbf{5.59}              &  0.75             &  6.28                     &  1.52    &  6.58                      &  0.66    &  7.05                         &  1.39    &  7.15                         &  1.46    \\ 
                                            &  173     &  \textbf{3.02}              &  0.18             &  3.11                     &  0.40    &  3.11                      &  0.23    &  3.23                         &  0.15    &  4.95                         &  1.86    \\ 
                                            &  384     &  1.98                       &  0.15             &  2.33                     &  0.33    &  \textbf{1.84}             &  0.09    &  1.87                         &  0.08    &  2.29                         &  0.15    \\
                                            &  480     &  1.75                       &  0.08             &  1.79                     &  0.10    &  1.78                      &  0.13    &  \textbf{1.58}                &  0.10    &  2.09                         &  0.14    \\ 
    \cline{1-12}
    \multirow{4}{*}{JavaGC}                 &  855     &  \textbf{15.99}             &  1.17             &  16.05                    &  0.95    &  20.79                     &  1.40    &  27.59                        &  1.02    &  20.41                        &  0.55    \\ 
                                            &  2571    &  \textbf{10.02}             &  0.44             &  11.17                    &  0.65    &  13.27                     &  0.57    &  18.61                        &  0.37    &  14.10                        &  0.33    \\ 
                                            &  3032    &  \textbf{9.74}              &  0.54             &  10.44                    &  0.79    &  12.24                     &  0.50    &  17.47                        &  0.39    &  13.92                        &  0.64    \\
                                            &  5312    &  \textbf{7.12}              &  0.24             &  8.66                     &  0.25    &  10.04                     &  0.38    &  14.85                        &  0.18    &  10.80                        &  0.24    \\ 
    \cline{1-12}
    \multirow{4}{*}{SaC}                    &  2060    &  \textbf{13.50}             &  0.94             &  15.59                    &  1.22    &  14.34                     &  0.75    &  22.49                        &  0.56    &  15.20                        &  0.92    \\ 
                                            &  2295    &  \textbf{12.94}             &  0.79             &  14.12                    &  1.06    &  14.51                     &  0.97    &  21.92                        &  0.61    &  13.86                        &  0.84    \\ 
                                            &  2499    &  \textbf{12.37}             &  0.70             &  14.10                    &  1.20    &  13.01                     &  0.65    &  20.57                        &  0.42    &  12.91                        &  0.59    \\
                                            &  3261    &  \textbf{11.48}             &  0.73             &  14.13                    &  0.97    &  12.54                     &  0.59    &  18.95                        &  0.43    &  11.68                        &  0.61    \\ 
    \cline{1-12}
  \end{tabular}}
\end{table*}

\subsection{RQ3: Hyperparameter Test} \label{subsec:hyper_test}

\subsubsection{Robustness of Regularization}

\begin{figure}[htbp]
  \centering
  \includegraphics[width=.95\columnwidth]{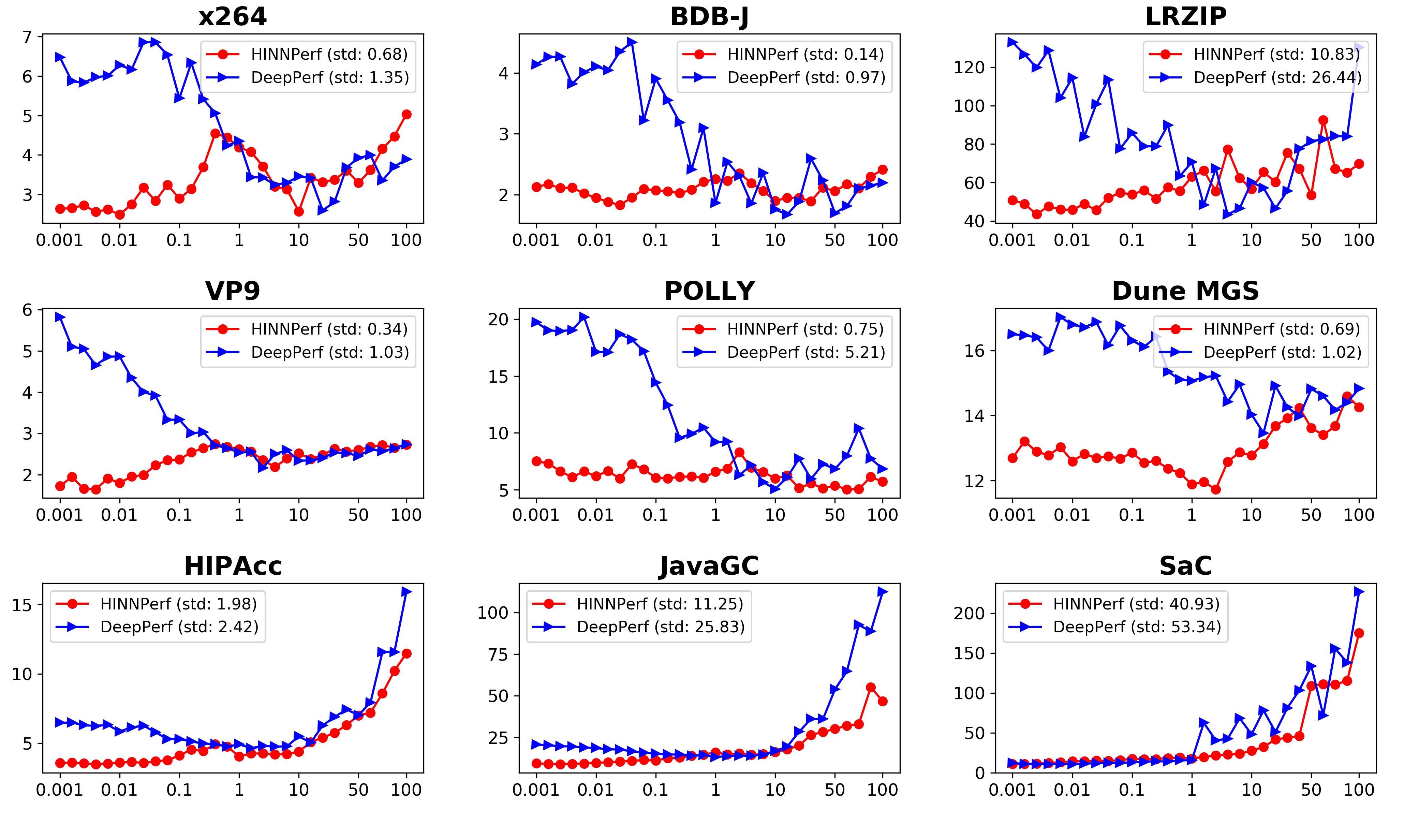}
  \caption{Results of hyperparameter sensitivity tests. Within each subfigure, the x-axis shows values of the regularization hyperparameter $\lambda$ while the y-axis shows the prediction MREs. And the std means the standard deviation of all points on the same polyline.}
  \label{fig:para_sen}
  \Description{regularization hyperparameter sensitivity tests}
\end{figure}

In the context of performance prediction with small sample, the $L_1$ regularization hyperparameter $\lambda$ of Equation~(\ref{equ:L_1_lambda}) is crucial for the model accuracy. Usually, it takes effort and time to search for the optimal hyperparameter value. \emph{DeepPerf} conducts grid search with 30 points logarithmically spaced in the range $[0.01,1000]$ to find the best value for the $L_1$ regularization hyperparameter~\cite{DeepPerf}. However, for the regularization hyperparameter of our proposed approach, we only use grid search with 5 points in $\{0.001, 0.01, 0.1, 1, 10\}$ to obtain better accuracy than \emph{DeepPerf}, reducing the search space by 6 times. We believe that the reduction of hyperparameter space benefits from the improvement of the model robustness. In this experiment, we demonstrate our hypothesis by testing the sensitivity of our model to the regularization hyperparameter $\lambda$ and compare the results with \emph{DeepPerf}. Figure~\ref{fig:para_sen} shows the prediction MREs of the two methods under different hyperparameter values logarithmically spaced in the range $[0.001,100]$. We present the results on 9 subject systems except the HSMGP system, since our method seems to make insignificant improvement on the HSMGP system from Table~\ref{tab:result}. We can observe that \emph{HINNPerf} is much less sensitive to the changes of regularization hyperparameter than \emph{DeepPerf} on most systems, demonstrating better robustness of our model. Hence, we conclude that the hierarchical $L_1$ regularization strategy makes \emph{HINNPerf} more robust to the regularization hyperparameter and thus we can search for the optimal hyperparameter value on a much smaller space.

\begin{figure}[htbp]
  \centering
  \includegraphics[width=.95\columnwidth]{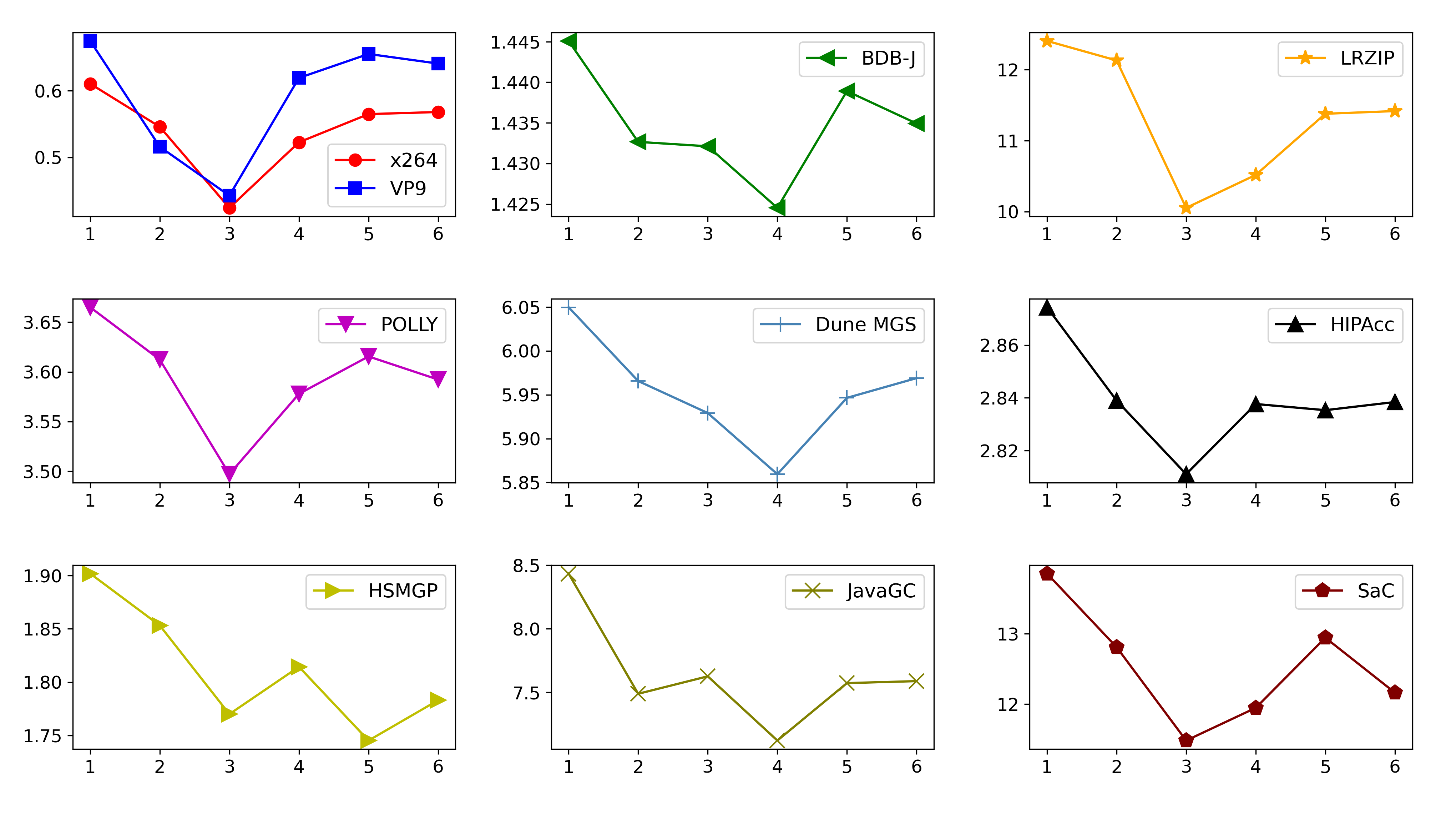}
  \caption{Prediction results of each system under different interaction orders. Within each subfigure, the x-axis shows values of the interaction order $m$ while the y-axis shows the prediction MREs.}
  \label{fig:m_sen}
  \Description{interaction hyperparameter tests}
\end{figure}

\subsubsection{Impact of Interaction Order}

In our experiment, \emph{HINNPerf} chooses the interaction order $m$ (i.e., the number of blocks) from $\{2,3,4,5\}$ and learns the best order value for each system from the data. We set the upper limit of $m$ to 5 degrees rigorously based on the findings of previous work~\cite{SPLConquer_02,SPLConquer_03,HierarchicalInter} that the interaction order of configurable systems is at most five. Moreover, it is necessary to investigate how different values of $m$ affect the efficacy of \emph{HINNPerf} on each system. Hence, we fix the hyperparameter $m$ to a default value of $1,2,3,4,5,6$ respectively and train the \emph{HINNPerf} to predict the performance of the ten subject systems. Here we additionally evaluate $m=6$ to see how the \emph{HINNPerf} behaves when $m>5$. Figure~\ref{fig:m_sen} presents the prediction errors of \emph{HINNPerf} under different interaction orders. It can be observed that each system has its own best interaction order and increasing or decreasing the optimal value of $m$ will reduce the performance prediction accuracy. We thus conclude that learning the interaction order $m$ from the data of each system is better than setting a default value manually. Besides, we find that setting $m=6$ does not cause significant differences\footnote{We have conducted the Wilcoxon rank sum test~\cite{ranksum} between the results of $m=5$ and $m=6$, and found that the $p$-value is bigger than 0.05.} with $m \le 5$, showing that it is possible for the \emph{HINNPerf} to learn the information of interaction order higher than $5$. However, to keep the consistency with previous work~\cite{SPLConquer_02,SPLConquer_03,HierarchicalInter}, we only consider $m \le 5$ in this paper and leave the exploration with $m>5$ to future work.

\subsection{RQ4: Model Practicality} \label{sec:interpretable}
\subsubsection{Coupling Complexity}

We leverage the hierarchical architecture of \emph{HINNPerf} to investigate whether the performance of the configurable systems is mainly affected by lower- or higher-order interactions, so as to provide additional insights into the complexity of coupling among configuration options. As discussed in Section~\ref{subsec:beyond_acc}, we first train the \emph{HINNPerf} model with the training dataset of a configurable system, and then predict the values of each partial performance function $f_{j}(\cdot)$ and the final output $f(\cdot)$ of Equation~(\ref{equ:perf_decom}) on the testing dataset. We measure the contribution of each partial function $f_{j}(\cdot)$ to the whole performance function $f(\cdot)$ through computing the average percentage: 
\begin{equation}
P_{f_j} = \frac{1}{|\mathcal{T}|} \sum_{\mathbf{o} \in \mathcal{T}} \frac{f_j(\mathbf{x}_j)}{f(\mathbf{o})}, ~~ 1 \le j \le m,
\end{equation}
where $\mathcal{T}$ denotes the testing dataset, $f(\mathbf{o})$ is the performance value of configuration $\mathbf{o}$, $f_j(\mathbf{x}_j)$ is the $j$-th partial performance value of configuration $\mathbf{o}$, $m$ is the number of subfunctions/blocks learned by \emph{HINNPerf}. As we describe in Section~\ref{subsec:theo_model}, different subfunctions $f_j(\cdot)$ learn the influences of different interaction orders. Therefore, the average percentage $P_{f_j}$ allows us to measure how much of the $j$-th order interaction affects the system performance.

\begin{figure}[t]
  \centering
  \includegraphics[width=.95\columnwidth]{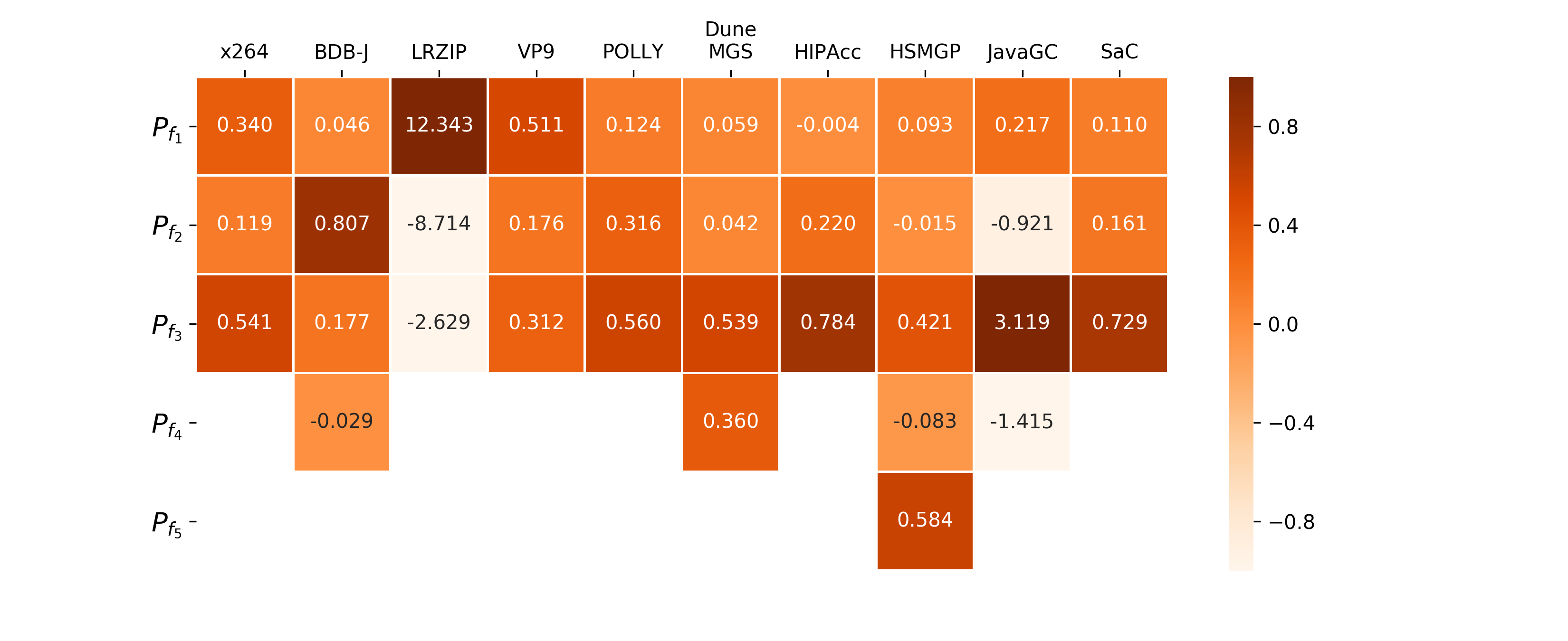}
  \caption{Contributions of different partial functions to the system performance. Note that different systems might have different numbers of partial functions (e.g., x264 has 3 partial functions while BDB-J has 4 partial functions) since the number of partial functions $m$ is a hyperparameter learned by \emph{HINNPerf}.}
  \label{fig:importance}
  \Description{contribution of different interactions}
\end{figure}

Figure~\ref{fig:importance} shows the contributions of different partial functions on the 10 subject systems, from which we obtain the following important observations:
\begin{itemize}
  \item Most configurable software systems have up to 3 partial performance functions with the second and the third subfunctions $f_2(\cdot), f_3(\cdot)$ having the most significant influence on the system performance. Since subfunctions $f_2(\cdot)$ and $f_3(\cdot)$ of our method aims to learn partial performance values of higher-order interactions, this observation implies that multi-way hierarchical interactions are common in configurable software systems and the most common interaction patterns might be two-way and three-way. These findings are consistent with previous work~\cite{SPLConquer_03} and demonstrate the feasibility of employing a hierarchical interaction architecture such as \emph{HINNPerf} to model the interactions among configuration options.
  \item For the system VP9 where the lower-order subfunction $f_1(\cdot)$ has the major influence on the system performance, \emph{HINNPerf} tells us that the configuration options of this system are relatively decoupled and have simple interactions. Accordingly, when using systems such as VP9, users can make moderately casual decisions on configuration selections without paying much attention to feature interactions. And developers can optimize each configuration option relatively separately to improve software performance.
  \item For the LRZIP system, however, the lower-order subfunction $f_1(\cdot)$ has a significant positive influence on the system performance while higher-order subfunctions $f_2(\cdot)$ and $f_3(\cdot)$ have negative influences, implying more complex relationships among configuration options. This may warn developers and users to carefully consider the feature interactions when deploying and using the system. Also, the JavaGC system should have similar concerns.
\end{itemize}

\begin{figure}[t]
\centering

\subfloat[$m = 3$]{ 
  \begin{minipage}[t]{0.5\columnwidth}
  \centering
  \includegraphics[width=\columnwidth]{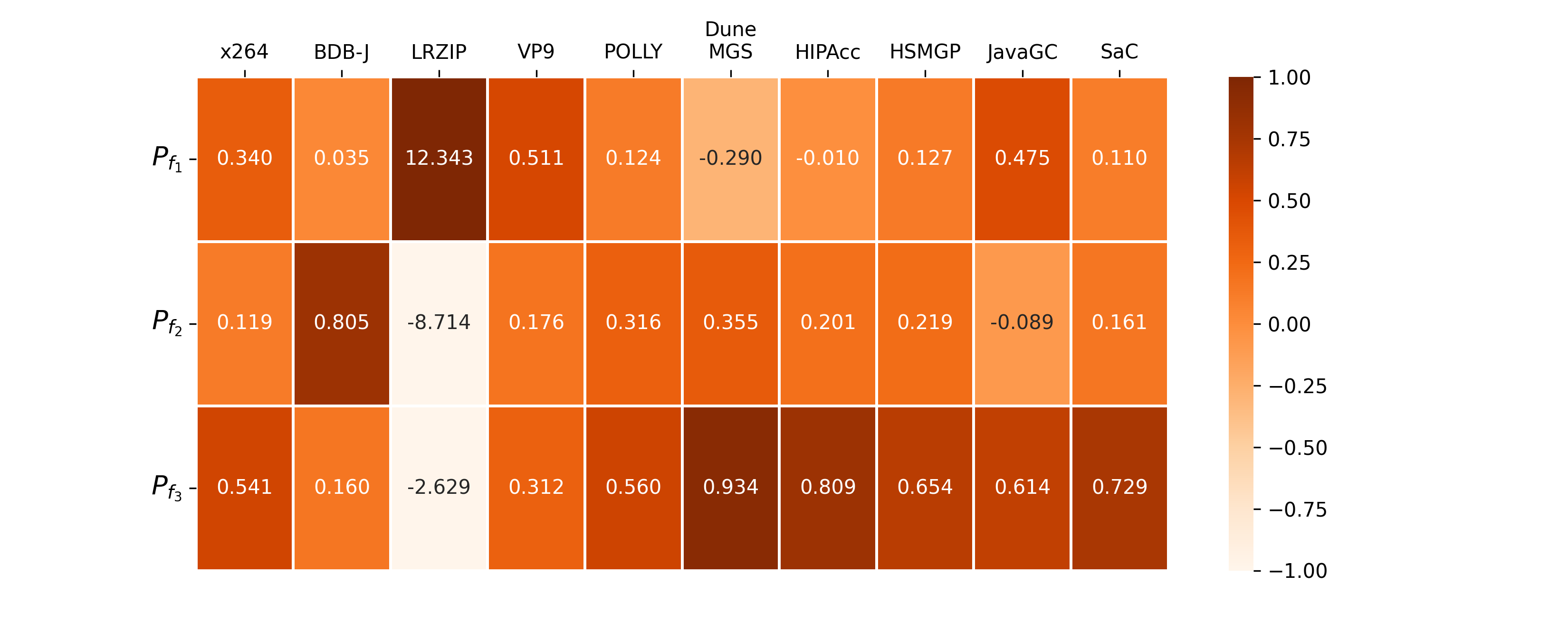}
  \label{fig:m3}
  \end{minipage}
}
\subfloat[$m = 4$]{ 
  \begin{minipage}[t]{0.5\columnwidth}
  \centering
  \includegraphics[width=\columnwidth]{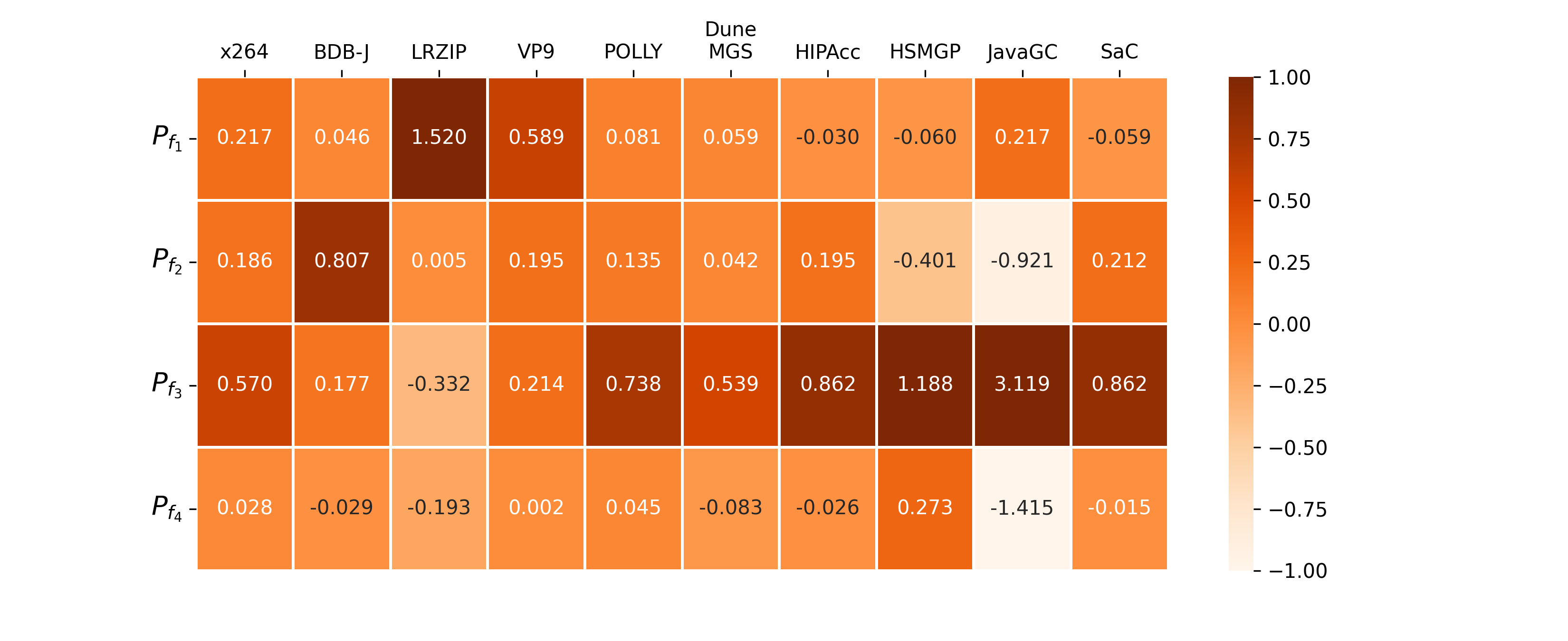}
  \label{fig:m4}
  \end{minipage}
}

\subfloat[$m = 5$]{ 
  \begin{minipage}[t]{0.5\columnwidth}
  \centering
  \includegraphics[width=\columnwidth]{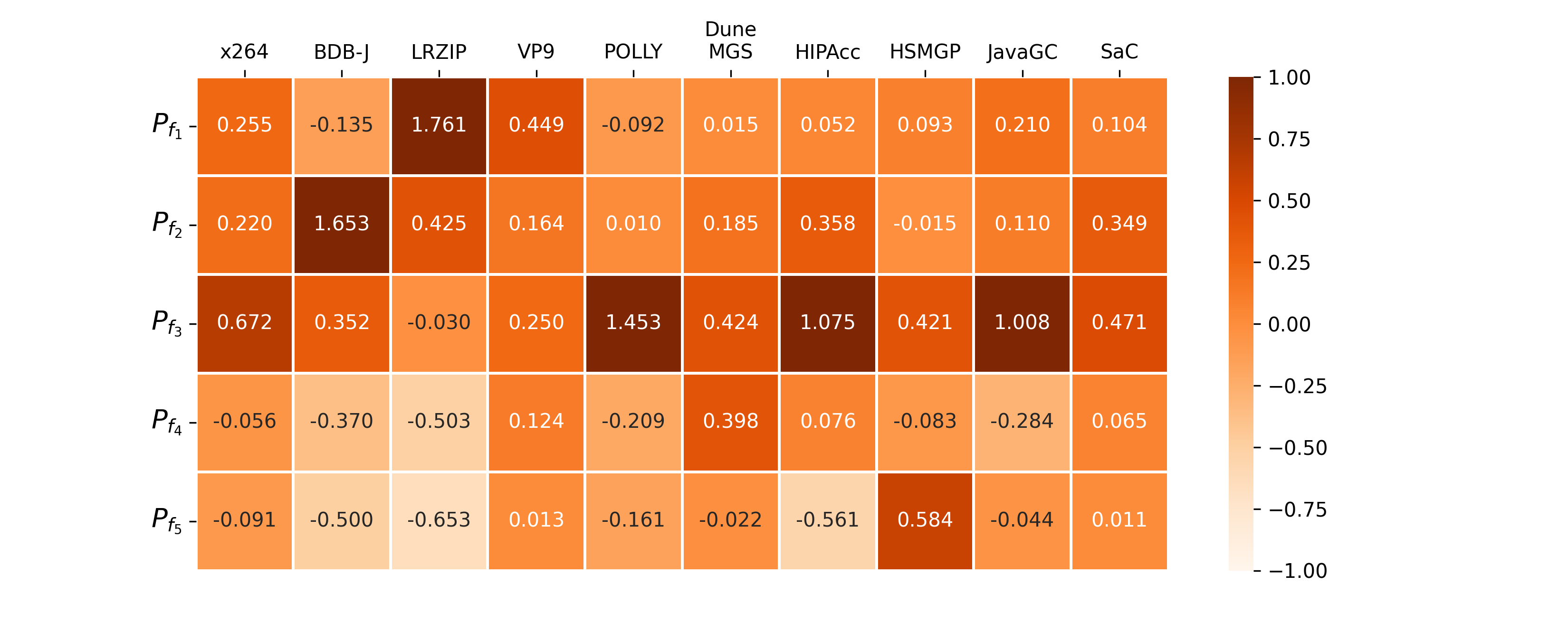}
  \label{fig:m5}
  \end{minipage}
}
\subfloat[$m = 6$]{ 
  \begin{minipage}[t]{0.5\columnwidth}
  \centering
  \includegraphics[width=\columnwidth]{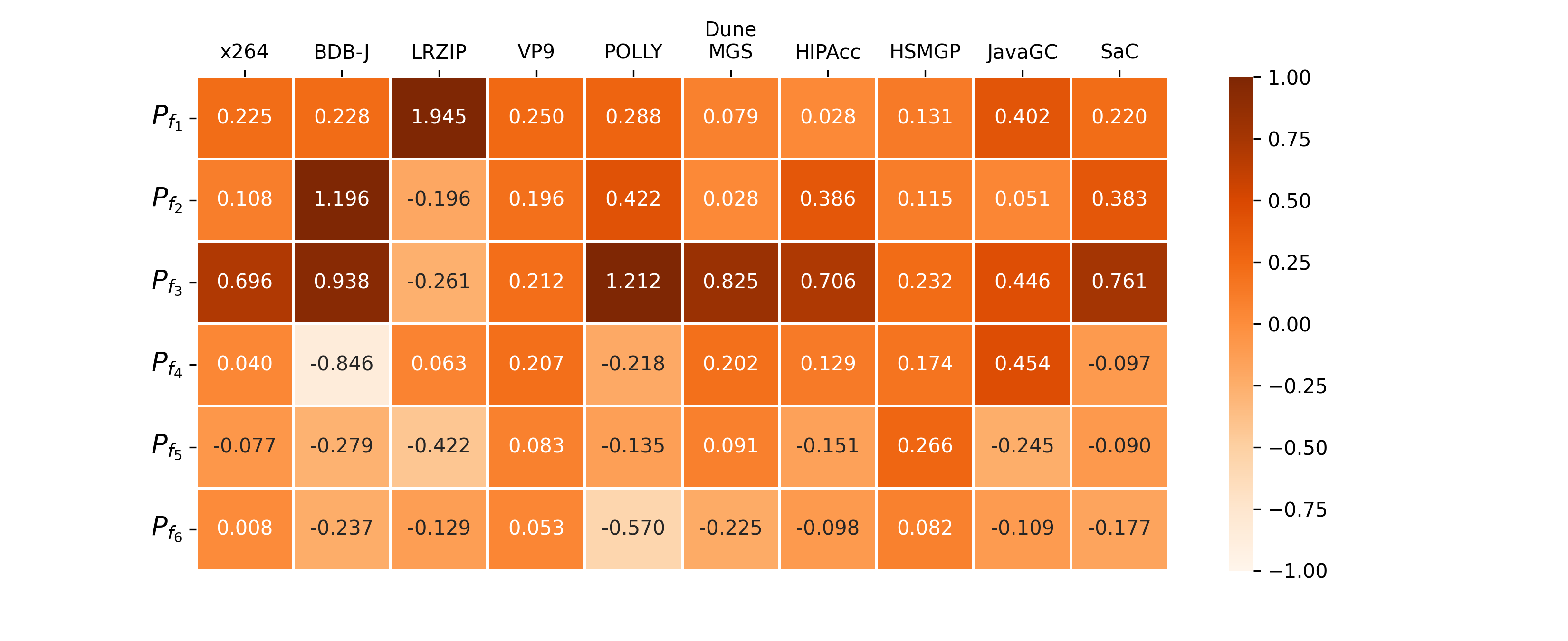}
  \label{fig:m6}
  \end{minipage}
}

\caption{Contributions of different partial functions to the system performance under $m=3,4,5,6$, respectively.}
\Description{contributions of different partial functions}
\label{fig:default_importance}
\end{figure}

The observations above demonstrate that our \emph{HINNPerf} model provides some insights about which kind of interactions have the most siginificant influence on the system performance. And we believe that these insights can help users and developers better understand the complexity of system coupling and make more rational decisions when deploying and using configurable systems. Note that \emph{HINNPerf} \emph{automatically} learns these insights from data without any manual intervention, which ensures the objectivity and the efficiency of our model.

While employing the \emph{HINNPerf} to automatically select the best value of $m$ for each system seems promising, we also interested in how \emph{HINNPerf} reveals the coupling complexity when a default value of $m$ is used. Besides, for the HSMGP system in Figure~\ref{fig:importance}, it could be worth testing for the six levels of interactions (i.e., $m=6$). Hence, we fix the hyperparameter $m$ to a default value of $3, 4, 5, 6$ respectively and train the HINNPerf to learn the contribution of each partial function. Figure~\ref{fig:default_importance} shows the results under different values of $m$, most of which are consistent with Figure~\ref{fig:importance}. For example, Figure~\ref{fig:importance} shows that for the x264, POLLY, Dune MGS, $\text{HIPA}^{cc}$, JavaGC and SaC systems, the third-order interaction (i.e., $P_{f_3}$) has the most significant influence on the system performance. This is also true in Figure~\ref{fig:default_importance}, no matter $m=3,4,5$ or $6$. And Figure~\ref{fig:m6} shows that the contribution of the $6$-th interaction order has insignificant contribution to the performance of all systems, including the HSMGP. Therefore, we believe that \emph{HINNPerf} can still capture the importance of each interaction order when a (proper) default value of $m$ is used.

\subsubsection{Option Significance}

\begin{figure}[t]
\centering

\subfloat[Option Significance of x264]{ 
  \begin{minipage}[t]{0.32\columnwidth}
  \centering
  \includegraphics[width=\columnwidth]{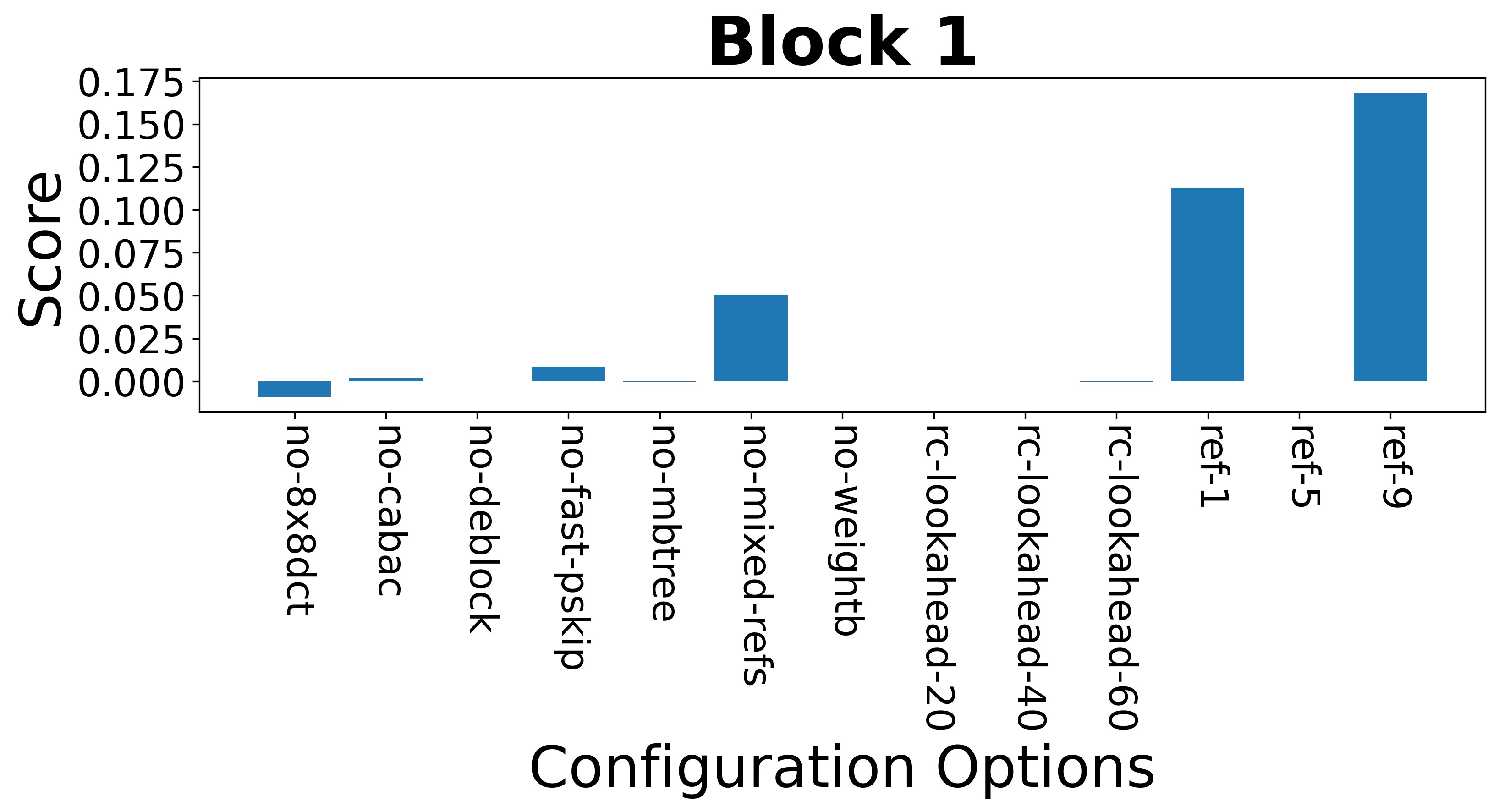}
  \label{fig:x264_1}
  \end{minipage}
  \begin{minipage}[t]{0.32\columnwidth}
  \centering
  \includegraphics[width=\columnwidth]{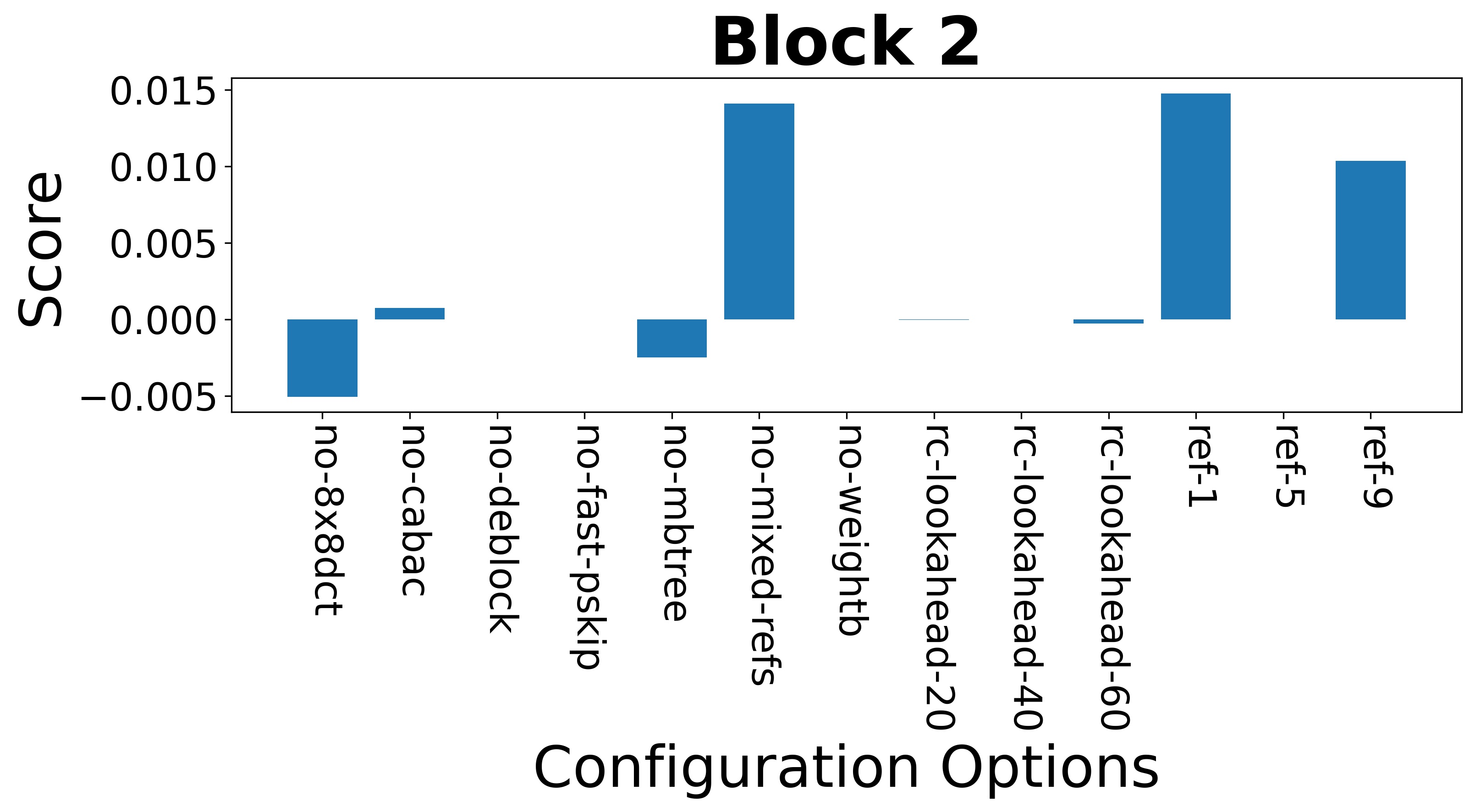}
  \label{fig:x264_2}
  \end{minipage}
  \begin{minipage}[t]{0.32\columnwidth}
  \centering
  \includegraphics[width=\columnwidth]{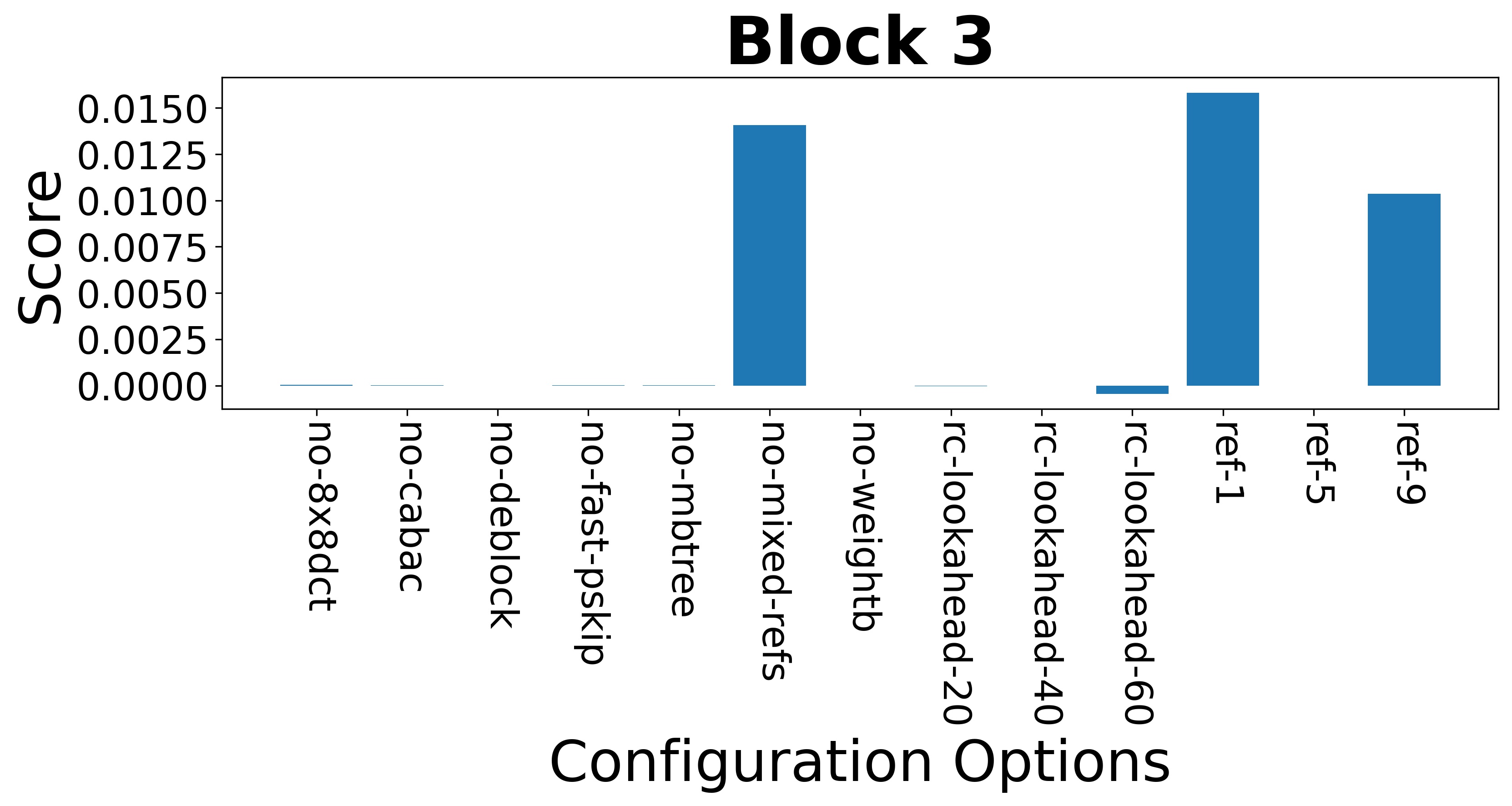}
  \label{fig:x264_3}
  \end{minipage}
}

\subfloat[Option Significance of SaC]{ 
  \begin{minipage}[t]{0.32\columnwidth}
  \centering
  \includegraphics[width=\columnwidth]{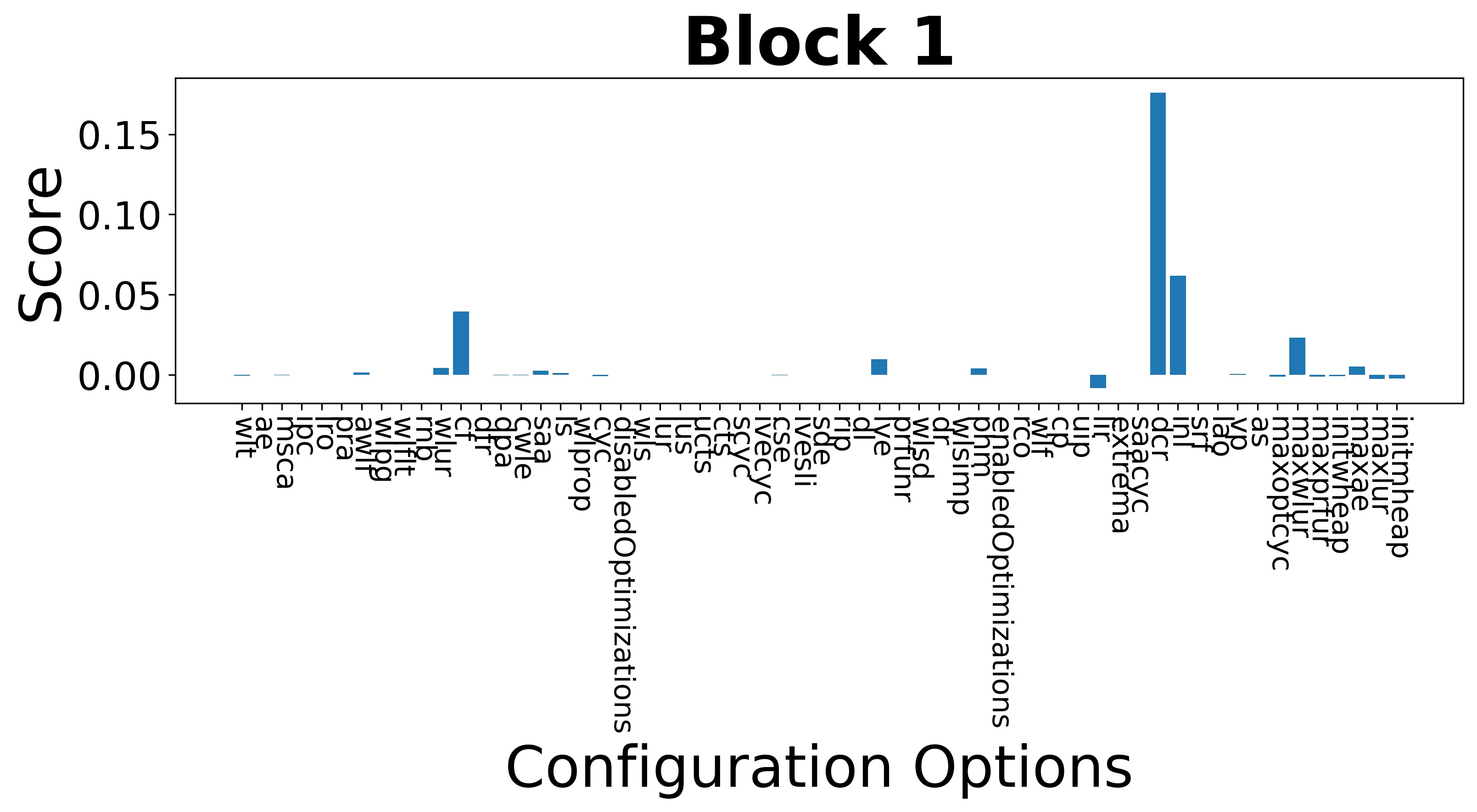}
  \label{fig:sac_1}
  \end{minipage}
  \begin{minipage}[t]{0.32\columnwidth}
  \centering
  \includegraphics[width=\columnwidth]{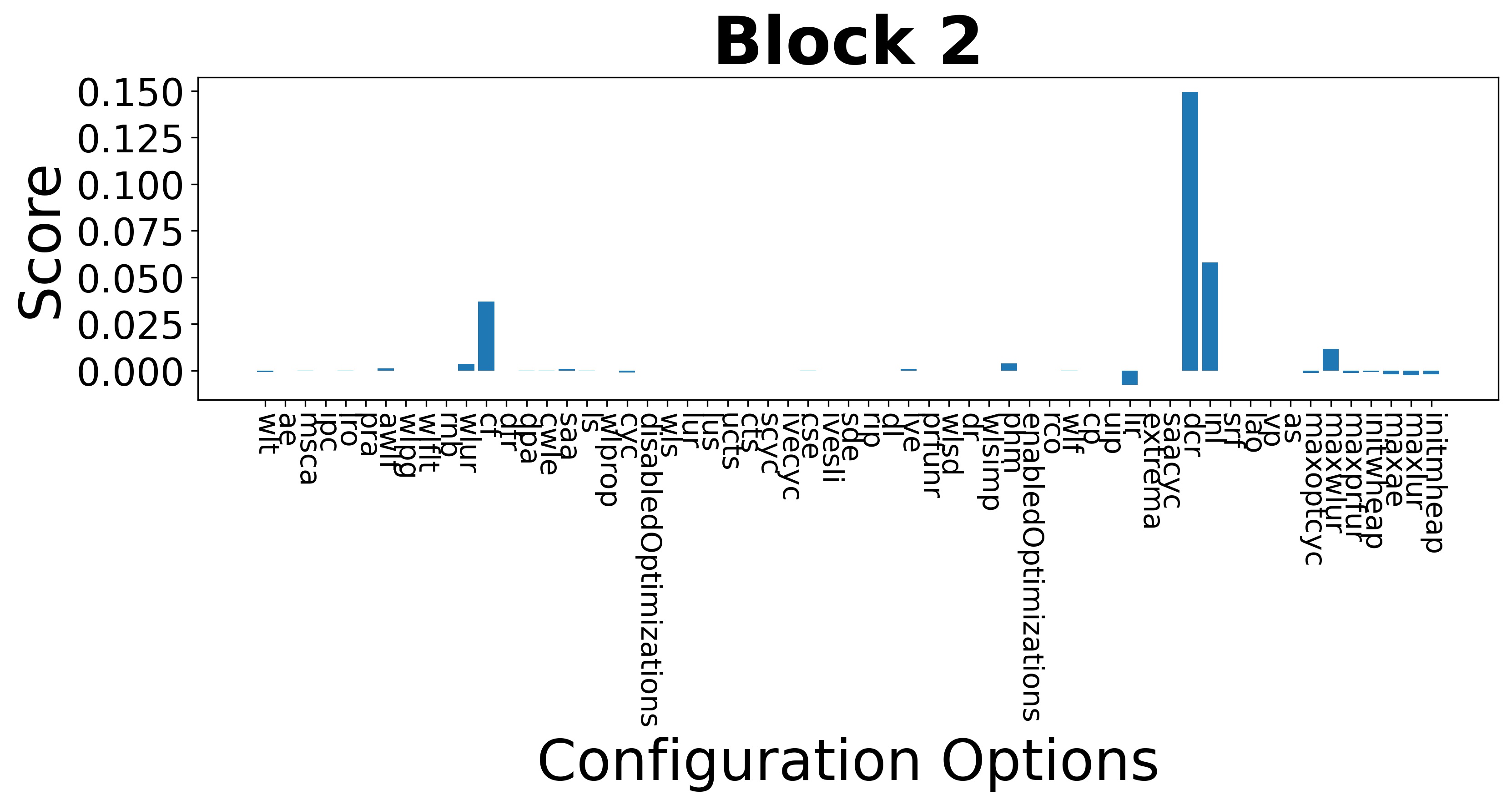}
  \label{fig:sac_2}
  \end{minipage}
  \begin{minipage}[t]{0.32\columnwidth}
  \centering
  \includegraphics[width=\columnwidth]{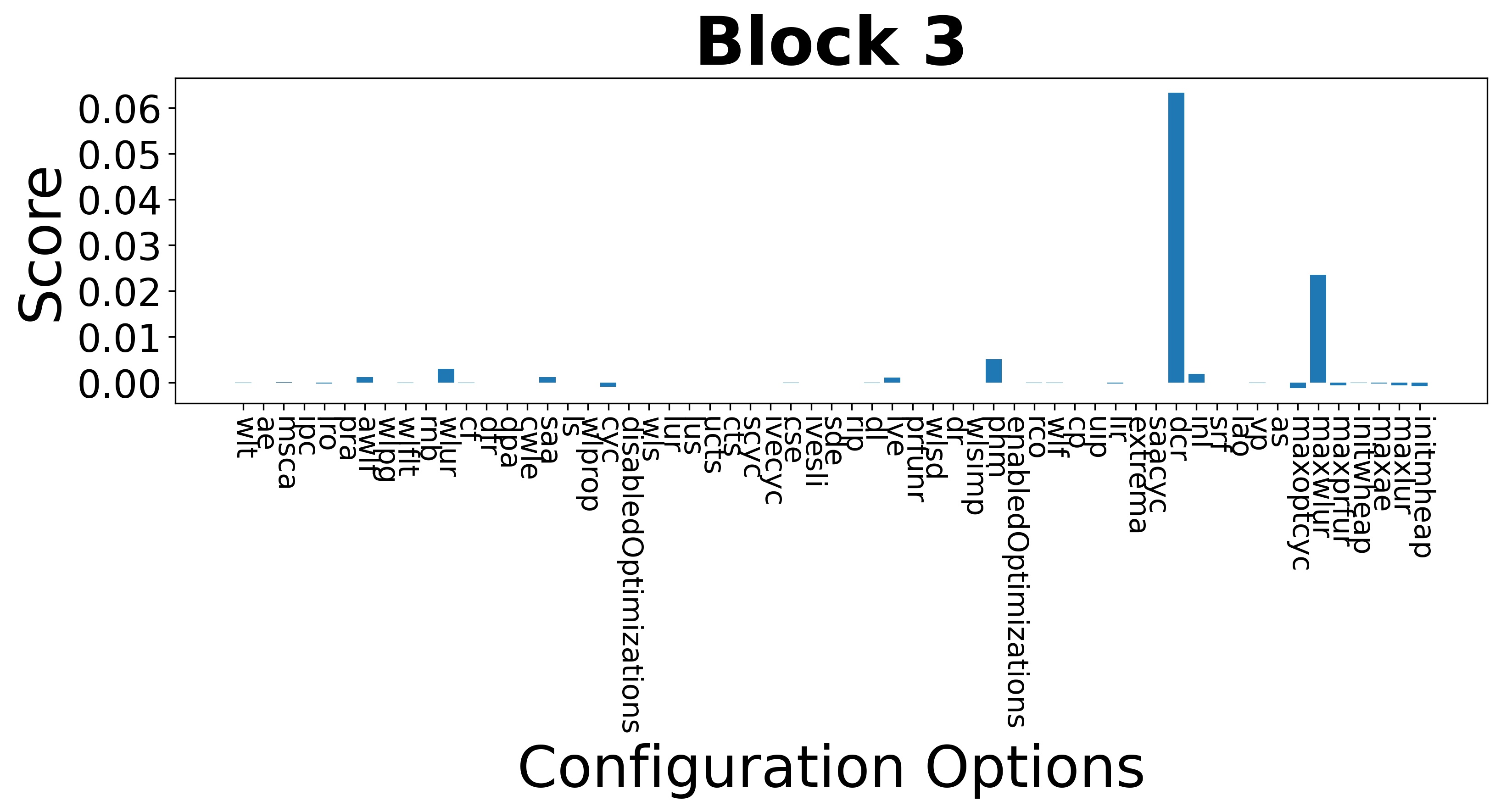}
  \label{fig:sac_3}
  \end{minipage}
}

\subfloat[Option Significance of $\text{HIPA}^{cc}$]{ 
  \begin{minipage}[t]{0.32\columnwidth}
  \centering
  \includegraphics[width=\columnwidth]{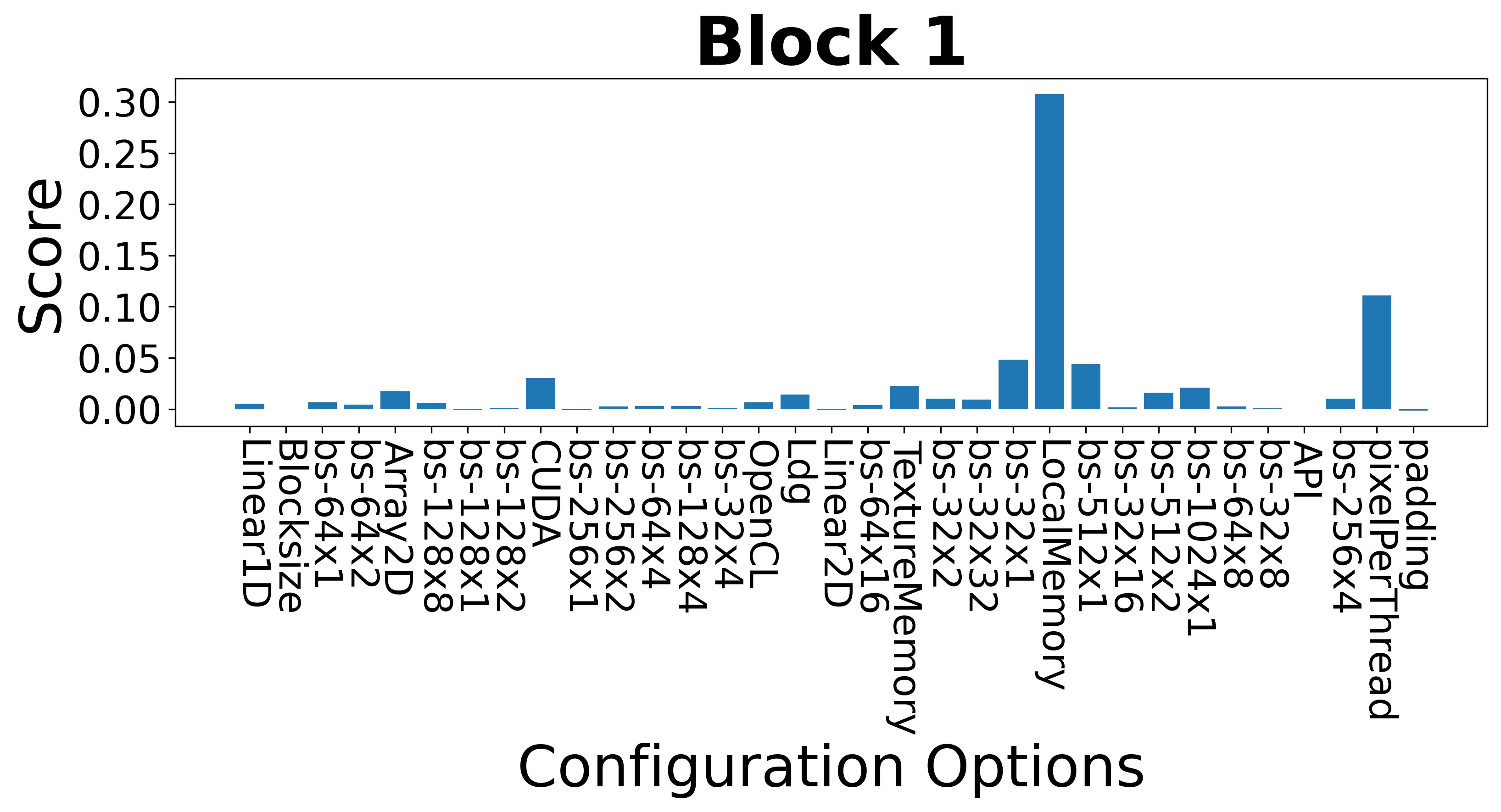}
  \label{fig:hipacc_1}
  \end{minipage}
  \begin{minipage}[t]{0.32\columnwidth}
  \centering
  \includegraphics[width=\columnwidth]{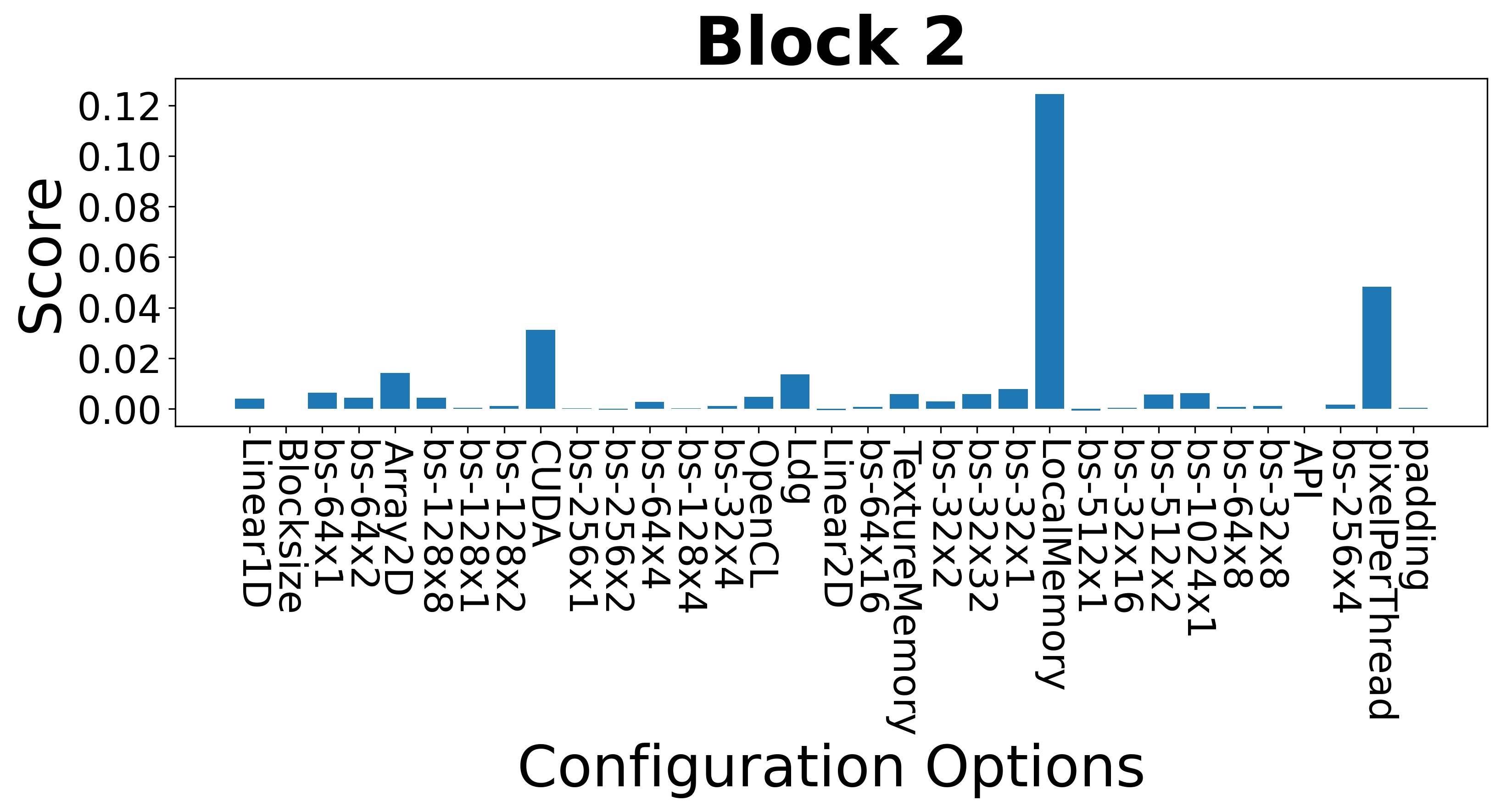}
  \label{fig:hipacc_2}
  \end{minipage}
  \begin{minipage}[t]{0.32\columnwidth}
  \centering
  \includegraphics[width=\columnwidth]{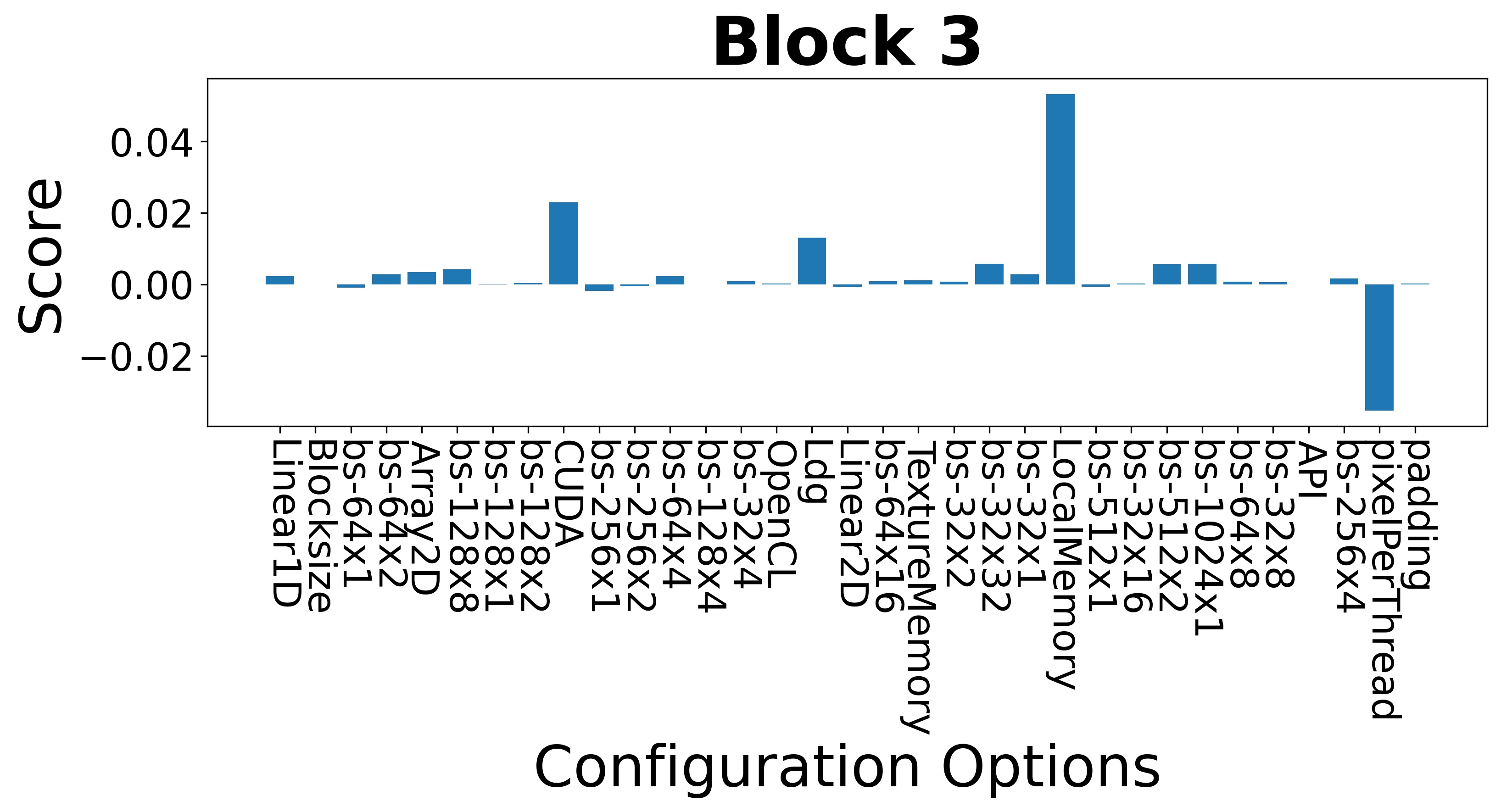}
  \label{fig:hipacc_3}
  \end{minipage}
}

\caption{The significance score of each configuration option in the three subject systems x264, SaC and $\text{HIPA}^{cc}$. Block $j$ denotes the $j$-th interaction order of the system, which corresponds to the subfunction $f_j(\cdot)$.}
\Description{significance scores}
\label{fig:sig_score}
\end{figure}

Based on previous insights about the percentage importance of different interaction orders, we further explore the significance of every configuration option in each interaction level. Recognizing the option significance in different interaction orders could help us pinpoint the options that have great impact on the performance and also provide some insights about the interaction patterns. Here we employ the Integrated Gradients~\cite{IG} method to assign a significance score to each configuration option:
\begin{equation}
\text{Score}(o_i) = (o_i - o_i') \int_{\alpha=0}^1 \frac{\partial f(\mathbf{o}' + \alpha (\mathbf{o} - \mathbf{o}'))}{\partial o_i} \mathrm{d}\alpha, ~~1 \le i \le n, 
\end{equation}
where $o_i$ is the $i$-th option of the configuration $\mathbf{o}$, $\mathbf{o}' \in \mathbb{R}^n$ is the baseline zero vector and $f(\cdot)$ denotes the performance function. We compute the significance score of each option $o_i$ in different blocks of \emph{HINNPerf} and report the option significance of three subject systems x264, SaC and $\text{HIPA}^{cc}$ in Figure~\ref{fig:sig_score} for examples. Comparing the option significance in different configurable systems, we can further extract some valuable information for both users and developers:
\begin{itemize}
  \item For most configurable software systems, only a small number of configurations and their interactions have significant impact on the performance. Figure~\ref{fig:sig_score} shows that there are only $3/16$, $4/60$ and $4/33$ significant configuration options in the x264, SaC and $\text{HIPA}^{cc}$ systems, respectively. This finding is also consistent with all the previous work~\cite{DeepPerf,SPLConquer_02,SPLConquer_03}. The main contribution of our work here is that we recognize the specific significant options through \emph{automatic learning} without any manual efforts. For example, we can identify that the performance of the x264 system is mainly affected by three significant options {\tt no-mixed-refs}, {\tt ref-1}, and {\tt ref-9} from Figure~\ref{fig:x264_1}.
  \item Combining with insights about coupling complexity in the previous section, we can quickly pinpoint configuration options that might have the greatest impact on the system performance. For instance, Figure~\ref{fig:importance} shows that the third-order interactions ($P_{f_3}$) contribute the most to the SaC system performance, which corresponds to Block~3 in Figure~\ref{fig:sac_1}. Accordingly, we could roughly infer that the most significant options {\tt dcr} and {\tt maxwlur} in Block~3 has the greatest influence on the performance of SaC, so as to remind users and developers to pay special attention to these two options when using the system.
  \item The option significance in different interaction orders also reveals some potential interaction patterns of the configurable system. For the $\text{HIPA}^{cc}$ system in Figure~\ref{fig:hipacc_1}, the {\tt LocalMemory} and the {\tt pixelPerThread} options have the greatest importance in the first-order interaction (Block 1). After entering into the second-order interaction (Block 2), the contribution of the {\tt CUDA} option increases. Finally in the third-order interaction (Block 3), the {\tt CUDA}, {\tt Ldg}, and {\tt LocalMemory} options maintain positive significance while the {\tt pixelPerThread} option changes into negative significance. This observation inspires us that the $\text{HIPA}^{cc}$ system might have interaction patterns in the following form:
  \begin{align}
  \begin{split}
  & I_p(\text{{\tt LocalMemory}},~ \text{{\tt pixelPerThread}}) + I_q(\text{{\tt CUDA}},~ \text{{\tt LocalMemory}}, ~\text{{\tt pixelPerThread}}) \\
  & + I_r(\text{{\tt CUDA}}, ~\text{{\tt Ldg}},  ~\text{{\tt LocalMemory}}, ~-\text{{\tt pixelPerThread}}),
  \end{split}
  \end{align}
  where $I_p, I_q, I_r$ could be arbitrary interaction functions. Such an interaction form is intuitive since the usage and communication of the local memory between CPU and GPU ({\tt CUDA}) can significantly affect the system performance, while the number of pixels calculated per thread is also highly related to the local memory capacity and the GPU computing power.
\end{itemize}

In summary, by inspecting the option significance in different interaction orders learned by \emph{HINNPerf}, one can efficiently locate the small subset of configuration options and their interactions that have the greatest impact on system performance, instead of wasting a huge amount of time searching hard through a large number of configurations. Besides, \emph{HINNPerf} can provide some insights on potential interaction patterns of the system through the hierarchical interaction architecture. We believe these advantages are more actionable for users and developers to deploy and test the configurable software systems.

\subsection{RQ5: Time Cost}

\begin{figure}[htbp]
  \centering
  \includegraphics[width=.75\columnwidth]{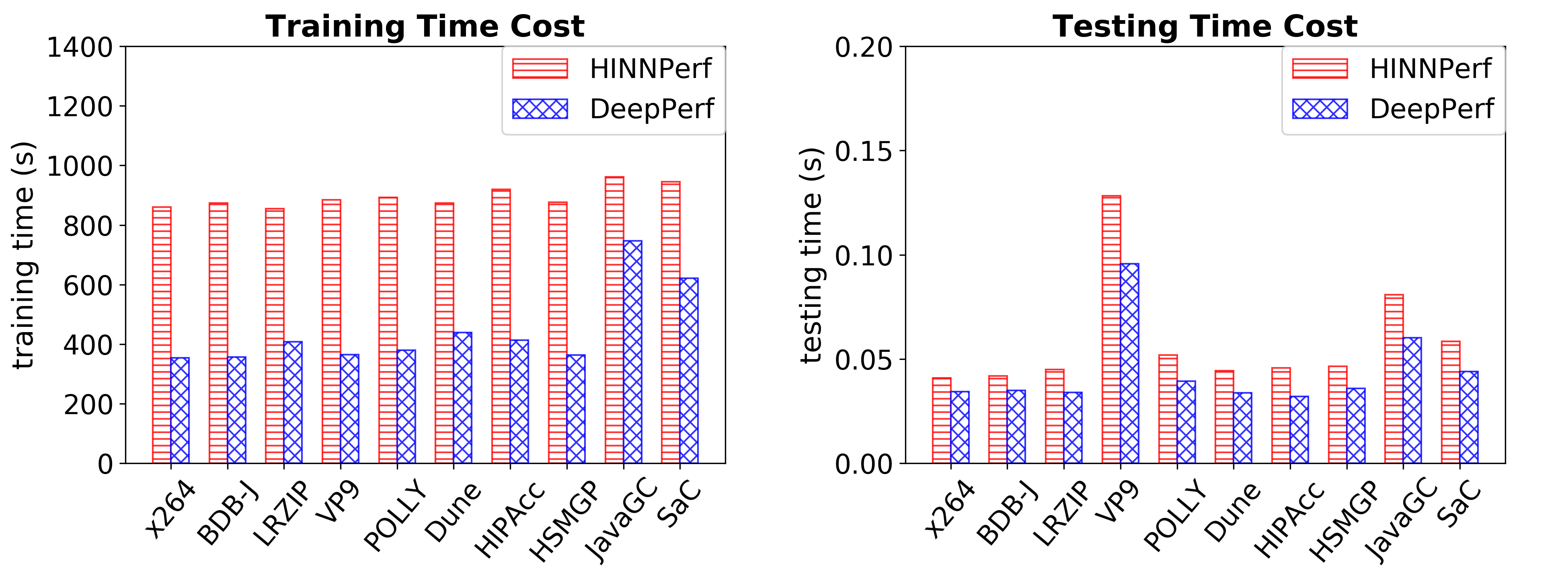}
  \caption{Time cost of the two deep learning methods on the 10 subject systems}
  \label{fig:time_cost}
  \Description{time cost}
\end{figure}

Although our method attains higher prediction accuracy and is more robust to the regularization hyperparameter than other approaches, it is also a more complex architecture with quantities of trainable parameters. Therefore, to evaluate the practicality and feasibility of our approach, it is necessary to measure the time consumed by the training and the testing process of \emph{HINNPerf}. Since we implement the two deep learning methods \emph{DeepPerf} and \emph{HINNPerf} under CPU and GPU environment while the other methods \emph{SPLConqueror}, \emph{DECART} and \emph{RF} under the environment without GPU, we only report the time cost of \emph{DeepPerf} and \emph{HINNPerf} on 10 subject systems with maximum sample size. As shown in Figure~\ref{fig:time_cost} (left), for all the systems with various configurable options, it takes \emph{HINNPerf} 14 - 16 minutes to do model training and hyperparameter searching, while \emph{DeepPerf} spends 5 - 13 minutes. Even though \emph{HINNPerf} takes longer time to train the performance model than \emph{DeepPerf}, the time cost of \emph{HINNPerf} is still acceptable. As for testing process, Figure~\ref{fig:time_cost} (right) shows that both methods perform similarly with time cost less than 0.2 seconds, which is consistent with our analysis in Section~\ref{subsec:complexity} that \emph{HINNPerf} only increases constant complexity than \emph{DeepPerf}.

In addition to model complexity, the main reason why the training of our method is more time-consuming than \emph{DeepPerf} is that \emph{DeepPerf} employs a more efficient hyperparameter search strategy. It first searches the optimal number of hidden layers for a non-regularized FNN. After applying the same number of hidden layers as the non-regularized FNN on the \emph{DeepPerf} architecture, it then searches for the best regularization hyperparameter value. Such a separation hyperparameter search strategy can significantly reduce the hyperparameter space, but it is not suitable for all deep neural network architectures (i.e., not universally applicable). What's worse, this strategy risks hurting model accuracy because the optimal hyperparameters of the non-regularized FNN may not be optimal for \emph{DeepPerf}. Instead, we directly search all optimal hyperparameters for the \emph{HINNPerf} architecture, slightly increasing the training time but guaranteeing high prediction accuracy. Besides, compared to the search strategy with accuracy drawdown risk, it is more reasonable to develop a robust network structure to reduce the hyperparameter space.

\subsection{Discussions}

\subsubsection{Strengths and Limitations}

The main strength of our \emph{HINNPerf} method is that it can model complicated interactions among the configuration options and attain higher prediction accuracy on various configurable systems than other state-of-the-art approaches. As shown in our experiments, for most of the software systems with binary and numeric options, compared to other baselines, \emph{HINNPerf} achieves higher prediction accuracy with less sample. For software systems with binary options, \emph{HINNPerf} can achieve better or comparable results compared to other methods.

\emph{HINNPerf}'s second strength is that it provides additional insights about the contributions of different interactions to the system performance and significant configuration options in different interation orders, which could reveal some potential interaction patterns of the configurable system and help users and developers efficiently make rational decisions on configurations. We emphasize that unlike the \emph{SPLConqueror} method~\cite{SPLConquer_03}, \emph{HINNPerf} learns these insights from data \emph{automatically} without any human efforts to design the interaction patterns. Another automated approach \emph{DeepPerf}~\cite{DeepPerf} does not show any additional insights in addition to performance prediction. Hence, compared to other performance models, our method is more efficient and helpful for users and developers.

Finally, the third strength of \emph{HINNPerf} is that it is less sensitive to the regularization hyperparameter and more robust than another deep learning method \emph{DeepPerf}~\cite{DeepPerf}. Therefore, our method effectively reduces the hyperparameter tuning cost and improves the training efficiency.

The major limitation of \emph{HINNPerf} is that it takes longer time to train than other baseline methods. However, considering the benefits of accuracy improvement and interaction insights brought by \emph{HINNPerf}, it is worth taking slightly more time to train a more useful performance model. Also, we will optimize the network structure and explore more efficient deep learning methods in the future work.

\subsubsection{Difference with DeepPerf}

Although our method is designed based on the feedforward neural network (FNN) as \emph{DeepPerf}, \emph{HINNPerf} is significantly different from \emph{DeepPerf} in the following aspects. First, \emph{HINNPerf} is a deep neural network model specially designed for performance prediction of configurable systems, which considers the unique characteristics of hierarchical interaction patterns in configurable systems. However, \emph{DeepPerf} is a general FNN model that has been widely used in deep learning field without specific design for the configurable systems. From this point of view, \emph{HINNPerf} makes an essential contribution to the field of deep learning-based performance prediction than \emph{DeepPerf}. Second, the hierarchical network architecture design allows \emph{HINNPerf} to recognize important options in different interaction orders and reveal some potential interaction patterns of configurable systems, which could help users and developers better understand how configurable systems work and efficiently identify significant options affecting the performance. This makes \emph{HINNPerf} more interpretable and practical than \emph{DeepPerf} which cannot provide any information about the internal interaction structure in configurable systems. We believe that the advantages guarantee the novelty and contribution of our method.

\subsubsection{Effectiveness of Neural Networks}

Although deep neural networks generally requires large data for training, the superiority of \emph{DeepPerf} and \emph{HINNPerf} over traditional machine learning (ML) methods such as \emph{SPLConqueror}, \emph{DECART} and \emph{RF} shows that deep neural networks are also effective in the performance prediction domain with small data. The main reason is probably that the excellent non-linear learning ability enables neural networks to better capture the complex interactions between different configuration options. Although traditional ML methods are more intuitive to process small data, the poor learning ability prevents them from learning the complicated interactions and achieving higher accuracy in the performance prediction task domain. Instead, as demonstrated in Section~\ref{sec:interpretable} (RQ4), the non-linear activation mechanism in the neural network allows it to extract richer information from the data to compensate for the shortcomings of the small sample. Therefore, it is necessary to employ deep neural networks to better learn the complex hierarchical interaction patterns in configurable systems and attain higher accuracy for performance prediction, which has been demonstrated by the experiment results in Section~\ref{sec: results} (RQ1).

\subsubsection{Threats to Validity}

For internal validity, to minimize the measurement bias of random sampling, we select samples of different sizes from each subject system to train all the methods, and then evaluate the prediction accuracy of each method on a test dataset. The test dataset excludes all the configurations of the training and validation datasets so as to evaluate the real predictive ability of each model. Besides, we repeat the random sampling, training and testing process 30 times and report the averages and the 95\% confidence intervals of the mean relative error for analysis. In this way, we believe that we control the measurement bias sufficiently and the experimental results reported in this paper are solid.

To increase external validity, we evaluate all the approaches on public datasets of 10 configurable systems. These systems are from different domains, with varying number of configuration options and different implementation languages. Although there is no guarantee that our method works for all other configurable systems, we are confident that the extensive experiments in this paper control this threat sufficiently.

The workload used to measure the configurable systems is also an important factor which can influence system performance but ignored by our work. The related studies~\cite{SPLConquer_02,SPLConquer_03,CART,DECART,DeepPerf,distance-sampling,PerLasso,Fourier} have not considered the workload factor either, which may be attributed to the great difficulty in quantifying the factor. According to previous work~\cite{SPLConquer_02,SPLConquer_03,distance-sampling}, all the datasets of the ten subject systems used in our experiments are collected under different workload settings, probably leading to mix workload in the data. However, in this work, we focus on designing a novel model for more accurate performance prediction. Thus, for fair comparison with the previous work, we directly adopt the widely-used \emph{public datasets} released by~\cite{SPLConquer_02,SPLConquer_03,distance-sampling}. With same datasets for comparison, the mix workload would not be an issue in our study, and the influence of workload on the performance prediction task is avoided. The experiment results in Table~\ref{tab:result} demonstrates that our method can achieve the state-of-the-art accuracy in different workload settings. In future work, we will explore the influence of the workload on performance modeling.

\section{Related Work}

Modern software systems are becoming increasingly complex and large-scale. Since poor performance can often be the cause of software project failure, how to manage system performance concerns along the software lifecycle has attracted great attention in the research and software industry communities~\cite{zibin_01,zibin_02,zibin_03,zibin_04}. For large-scale configurable systems, understanding the influences of the configuration options and their interactions on system performance plays an important role in software testing and maintenance phases~\cite{PerLasso}.

Recently, software researchers have conducted a large amount of work on performance prediction of highly configurable systems. In Section~\ref{sec:Intro}, we have discussed pros and cons of most state-of-the-art approaches, including \emph{SPLConqueror}~\cite{SPLConquer_01,SPLConquer_02,SPLConquer_03}, \emph{CART/DECART}~\cite{CART,DECART}, Fourier methods~\cite{Fourier,PerLasso}, and \emph{DeepPerf}~\cite{DeepPerf}. Throughout the history of performance models, the evolution of prediction methods tends to sacrifice the interpretability of the model to improve the prediction accuracy. \emph{SPLConqueror}~\cite{SPLConquer_01,SPLConquer_02,SPLConquer_03} is an early linear regression model that achieves low accuracy but can explicitly show the influences of configuration options and their interactions through regression coefficients. Later, \emph{CART/DECART}~\cite{CART,DECART} employs decision trees to improve the prediction accuracy for binary configurable systems, but can only qualitatively measures the feature importance through the tree structure. Recently, \emph{DeepPerf}~\cite{DeepPerf} utilizes deep feedforward neural networks to greatly improve the performance prediction accuracy, but with the model interpretability completely lost. On the other hand, the major contribution of Fourier learning~\cite{Fourier} is to derive a sample size that guarantees a theoretical boundary of the prediction accuracy. However, it attains no significant improvement in model interpretability and accuracy compared to \emph{SPLConqueror} and \emph{CART/DECART}, which is why we do not compare our model with Fourier methods in this paper.

Although our proposed \emph{HINNPerf} approach mainly focuses on achieving higher accuracy with less sample for performance prediction of configurable systems, we also make some progress in model interpretability by introducing the hierarchical interaction network architecture. In this way, users and developers can better understand how different interactions influence the system performance and make rational configuration decisions to maximize software efficacy. Besides, there is also some work on increasing explainability of the performance model by measuring the uncertainty in performance estimations with Bayesian method~\cite{mastering}. We will also consider the uncertainty measurement of our method in the future work.

Another related research topic is to select an optimal sample of configurations. Sayyad~et al.~\cite{SPL-selection-01,SPL-selection-02} propose evolutionary methods to select optimal features under multiple objectives. Later, Christopher~et al.~\cite{multiobj-sam} extend the multi-objective search-based optimization with constraint solver and improve Software Product Line (SPL) feature selection. Other researchers focus on selecting proper configuration option under certain coverage criterion~\cite{sample-criterion-01,sample-criterion-02,sample-criterion-03}. Recently, Christian~et al.~\cite{distance-sampling} propose a distance-based sampling strategy to better cover different kinds of interactions among configuration options in the sample set. Since we aim to propose a novel learning method for performance prediction in this paper, our method only employs random sampling strategy to build a performance model. In fact, we can combine our approach with different sampling work to further improve the model accuracy.

\section{Conclusion}

In this paper, we propose \emph{HINNPerf}, a novel hierarchical interaction neural network architecture for performance prediction of highly configurable systems. Our method decomposes the whole deep neural network into multiple hierarchical blocks and employs the embedding method to model the complex interactions among configuration options. Besides, we devise a hierarchical regularization strategy to ensure the model accuracy when trained on a small sample. The experimental results on public datasets show that \emph{HINNPerf} can better learn the complicated interactions in configurable systems and achieve better performance prediction accuracy with less data, when compared to other state-of-the-art methods. Furthermore, our approach provides additional insights for users and developers and is more robust than the advanced deep learning method \emph{DeepPerf}~\cite{DeepPerf}.

For future research, the proposed model can be further improved by incorporating effective sampling heuristics~\cite{SPLConquer_03,distance-sampling}. We will also explore the uncertainty measurement methods~\cite{mastering} to increase the interpretability of our approach.

\begin{acks}
The research is supported by the Key-Area Research and Development Program of Guangdong Province (2020B010165003), the National Natural Science Foundation of China under project (62032025, 62002084), Stable support plan for colleges and universities in Shenzhen under project (GXWD20201230155427003-20200730101839009). The corresponding author is Zibin Zheng.
\end{acks}

\bibliographystyle{ACM-Reference-Format}
\bibliography{TOSEM-2021-0156}

\end{document}